\documentclass[useAMS,usenatbib]{mn2e}
\usepackage{graphicx}
\usepackage{lscape}
\usepackage{multirow}
\usepackage{color}

\title[Automated Morphological Classification of SDSS Red Sequence Galaxies]{Automated Morphological Classification of SDSS Red Sequence Galaxies}
\author[Cheng et al.]{Judy Y. Cheng,$^{1}$\thanks{E-mail: jyc@ucolick.org} S. M. Faber,$^{1}$ Luc Simard,$^{2}$ Genevieve J. Graves,$^{3,4}$ \newauthor Eric D. Lopez,$^{1}$ Renbin Yan$^{5}$ and Michael C. Cooper$^{6,7}$\\
$^{1}$UCO/Lick Observatory, Department of Astronomy and Astrophysics, University of California, Santa Cruz, CA 95064 USA\\
$^{2}$National Research Council of Canada, Herzberg Institute of Astrophysics, Victoria, British Columbia, Canada\\
$^{3}$Department of Astronomy, University of California, Berkeley, CA 94720 USA\\
$^{4}$Miller Fellow\\
$^{5}$Department of Astronomy and Astrophysics, University of Toronto, Toronto, ON M5S 3H4, Canada\\
$^{6}$Steward Observatory, University of Arizona, Tucson, AZ 95721 USA\\
$^{7}$Spitzer Fellow}
\begin{document}

\date{Accepted xxxx Month xx. Received xxxx Month xx; in original form xxxx Month xx}

\pagerange{\pageref{firstpage}--\pageref{lastpage}} \pubyear{2009}

\maketitle

\begin{abstract}
In the last decade, the advent of enormous galaxy surveys has motivated the development of automated morphological classification schemes to deal with large data volumes. Existing automated schemes can successfully distinguish between early and late type galaxies and identify merger candidates, but are inadequate for studying detailed morphologies of red sequence galaxies. To fill this need, we present a new automated classification scheme that focuses on making finer distinctions between early types roughly corresponding to Hubble types E, S0, and Sa. We visually classify a sample of 984 non-starforming SDSS galaxies with apparent sizes $>14''$.  We then develop an automated method to closely reproduce the visual classifications, which both provides a check on the visual results and makes it possible to extend morphological analysis to much larger samples.  We visually classify the galaxies into three \textit{bulge classes} ($BC$) by the shape of the light profile in the outer regions: discs have sharp edges and bulges do not, while some galaxies are intermediate. We separately identify galaxies with features: spiral arms, bars, clumps, rings, and dust. We find general agreement between $BC$ and the bulge fraction $B/T$ measured by the galaxy modeling package GIM2D, but many visual discs have $B/T>0.5$. Three additional automated parameters -- smoothness, axis ratio, and concentration -- can identify many of these high-$B/T$ discs to yield automated classifications that agree $\sim70$ per cent with the visual classifications ($>90$ per cent within one $BC$). Tests versus disc inclination indicate that both methods identify most face-on discs, but visually, features are lost in edge-on discs. Eighty per cent of face-on visual discs have features while few visual bulges do, strongly validating the visual classifications. Given the good agreement between the visual and automated methods, we believe that the automated method can be applied to a much larger sample with confidence. Both methods are used to study the bulge vs. disc frequency as a function of four measures of galaxy \textquoteleft size': luminosity, stellar mass, velocity dispersion ($\sigma$), and radius ($R$). All size indicators show a fall in disc fraction and a rise in bulge fraction among larger galaxies.
\end{abstract}

\begin{keywords}
galaxies: evolution -- galaxies: structure -- galaxies: elliptical and lenticular -- galaxies: bulges
\end{keywords}

\section{Introduction}\label{intro}
With large-scale surveys like the Sloan Digital Sky Survey (SDSS, \citealp{yor00}), studies of galaxy properties are now possible for larger samples than ever before. The SDSS has obtained imaging of 11,663 square degrees of sky in five bandpasses and spectra of 929,555 galaxies in the local universe. Photometry and structural parameters are available for 357 million unique objects, and analyses of large samples of both spectra and photometry are the topic of current research. In the coming years, other large scale surveys, including Pan-STARRS\footnote{\footnotesize{http://pan-starrs.ifa.hawaii.edu/public/home.html}} and LSST,\footnote{\footnotesize{http://www.lsst.org/lsst}} will be adding to the available data.

The present study is motivated by the desire to bring to bear this great wealth of information on the question of the star formation histories of galaxies, and in particular how some galaxies had their star formation shut down such that their star formation rates today are low or zero; morphologically most of these are Es and S0s, hereafter early type galaxies. \citet{str01}, in an analysis of 150,000 SDSS galaxies, noted that the color distribution of all galaxies is distinctly \textit{bimodal}. At bright magnitudes, galaxies populate both a red sequence and blue cloud, which are separated by a dearth of galaxies in a green valley (e.g., \citealp{bal04}). On average, galaxies on the red sequence live in denser environments compared to blue galaxies (e.g., \citealp{hog03,bla05b,coo06,mar06,ski09}). Lookback observations have further shown that the relative fraction of red sequence galaxies is increasing with time, having at least doubled since $z=1$ \citep{bel04,bro07,fab07}.

Red sequence galaxies primarily owe their colors to a lack of hot, young stars, which are indicators of recent star formation. One persistent question in galaxy evolution is what mechanism causes a blue, star forming galaxy to \textit{quench} -- to stop forming new stars and become \textquoteleft red and dead' -- and migrate from the blue cloud to the red sequence. \citet{fab07} favored a \textquoteleft mixed scenario' for the formation of red galaxies to explain both the observed luminosity functions and the properties of nearby red galaxies. In this scenario, blue galaxies increase their mass through star formation and mergers until quenching occurs. The end products of this process populate the intermediate and less massive end of the red sequence. Massive ellipticals on the bright end of the red sequence are built up through dry (gas poor) mergers of these less massive galaxies along the red sequence.

This picture, however, still does not specify the exact physical mechanism(s) responsible for cutting off a galaxy's gas supply, thereby preventing the formation of new stars. At the risk of oversimplifying, major theories broadly fall into two classes. The first class says that feedback from the formation of a black hole or a starburst during a major merger can expel the gas (e.g., \citealp{san88,hop06}). Observations indicate that AGN are capable of releasing large amounts of energy (e.g., \citealp{fab06,raf06}), and simulations suggest that the incorporation of feedback mechanisms can reproduce the observed luminosity functions and colors of galaxies (e.g., \citealp{cro06}). 

The second class posits a critical dark halo mass ($M_{\rm crit}\sim10^{12}M_{\odot}$), above which cooling is too inefficient to allow central galaxies to accrete new gas (e.g., \citealp{ree77,blu84,dek06}). In this latter scenario, a satellite galaxy falling into such a large halo will also be quenched. Some recent observations have shown evidence of the existence of a critical halo mass (e.g., \citealp{bro08,pas09}).

It is surprisingly hard to distinguish the two theories observationally. For example, both pictures predict that most galaxies now undergoing quenching will be found in small groups, and thus in modestly overdense environments. The moderate velocity dispersions of such groups promote merging \citep{bin82}, as needed by the merger/feedback model, while the halo masses of such groups are also typically near $10^{12}M_{\odot}$, which puts them near $M_{\rm crit}$. Indeed, semi-analytic models indicate that many galaxies near $M^{*}$ in the mass function both experience major mergers and cross the $M_{\rm crit}$ halo boundary nearly simultaneously \citep{cat08}.

Galaxy morphologies, however, may be able to provide key independent data with which to test the two scenarios. Ellipticals, for example, are believed to be formed through major mergers \citep{too72}. Disc-dominated galaxies, on the other hand, cannot have undergone recent major mergers, as such mergers would have severely disrupted their shapes \citep{tot92}. Thus, neither class of quenching models is independently sufficient to explain the full range of morphological types observed for quenched galaxies. Mergers alone cannot explain the presence of quenched discs, and, likewise, a critical halo mass alone cannot explain why quenched galaxies typically have elliptical morphologies. A recent paper by \citet{van09}, for example, uses this latter point to argue that the roundness of galaxies with $M>10^{11}M_{\odot}$ is evidence for the importance of major merging in the formation of the most massive galaxies.

These two scenarios, of course, are not necessarily mutually exclusive, but if different mechanisms are responsible for the formation of bulges and discs (Es and S0s), then it is important to study them separately. In particular, the frequencies of each type may shed light on the relative importance of each mechanism. Thus, the bulge/disc frequency along the red sequence and versus other galaxy properties, such as galaxy size, will place an important new constraint on the theories of quenching.

Currently, only a rough knowledge of bulge/disc frequencies exists. Many groups have examined how the luminosity functions of different morphological types evolve with redshift (e.g., \citealp{bri00,bun05,fra06,ilb06,pan06,pan09}), but these studies lump Es and S0s together into one group. Other studies of nearby galaxies treat Es and S0s separately but have total sample sizes (including both early and late types) of only a few thousand (e.g., \citealp{mar94,mar99}). To quantify bulge/disc frequencies and study them in detail, we need a large sample of red sequence galaxies, such as is available from the SDSS, plus a method to classify them into bulges and discs.

One crucial requirement of such a method is the ability to identify star-forming galaxies that are on the red sequence because they are reddened by dust. For the SDSS Main Galaxy Sample \citep{str02}, this can be accomplished using spectroscopic information. In addition, these galaxies typically have features that are associated with star formation, such as spiral arms, clumps, or dust lanes, that distinguish them from galaxies that are truly quenched.

Red sequence galaxies, then, can be thought of as falling into three main groups: bulges, smooth discs, and unsmooth discs. This last group consists of discs with features generally associated with star formation (such as spiral arms, clumps, and dust) that cause the light profile of the galaxy to deviate from a smooth model. Neglecting bars and rings, which may be found in any disc-dominated galaxy, these three groups correspond roughly to the Hubble types E, S0, and Sa+later, respectively. These divisions accomplish our goal of distinguishing between bulges and discs, as well as accounting for the contamination of the red sequence by galaxies that may not be truly quenched (unsmooth discs).

Traditionally, morphological classification has been done by eye (e.g., \citealp{dev91,fuk07}). Visual classification, however, is time-consuming and the resulting galaxy samples are small -- two of the largest such samples consist of 2253 galaxies from SDSS \citep{fuk07} and 3314 galaxies from the Millennium Galaxy Catalog \citep{dri06}. One novel approach taken by the Galaxy Zoo team involves employing $\sim10^{5}$ \textquoteleft citizen scientists' to visually classify 40 million galaxies in the SDSS \citep{lin08}. At this time, however, their classifications make no distinction between Es and S0s.  To take advantage of the full sample of SDSS, we will need an \textit{automated} method of morphological classification that can be done quickly on a large sample of galaxies and is yet capable of making fine distinctions among early type galaxies.

Several automated methods, including the CAS system \citep{con03} and the Gini and $M_{20}$ parameters \citep{abr03,lot04}, have already been used to classify SDSS galaxies and are able to reliably distinguish between early and late type galaxies. But under these methods, the morphologies we are interested in distinguishing (bulges, smooth discs, unsmooth discs) are mostly grouped together as \textquoteleft early types.' Our approach is to capitalize on two existing sources of automated parameters, the SDSS photometric pipeline and the galaxy modeling package Galaxy IMage 2D (GIM2D; \citealp{sim02}) Galaxy IMage 2D (GIM2D; \citealp{sim02}), to find combinations of automated parameters that can successfully sort red sequence galaxies into bulges, smooth discs, and unsmooth discs.

For SDSS galaxies, one widely used automated parameter is the concentration $C=R_{90}/R_{50}$, where $R_{90}$ and $R_{50}$ are the radii containing 90 and 50 per cent of the Petrosian flux in the $r$-band, respectively.\footnote{\footnotesize{The Petrosian flux is defined as the flux contained within twice the Petrosian radius, the circular radius at which the local surface brightness $\mu(r)$ is equal to 20 per cent of the enclosed mean surface brightness $\mu(<r)$ \citep{bla01}.}} Classical bulges, which have bright central regions and extended outer envelopes, have high $C$ because the bulk of their light is located at small radii. \citet{shi01} and \citet{str01} find that dividing galaxies using $C$ gives automated samples of early and late types with about 15-20 per cent contamination from the opposite class. This is promising but still does not accomplish our goal of distinguishing Es and S0s.

The galaxy modeling package GIM2D has also been used to classify galaxies by measuring several quantitative parameters. GIM2D is an IRAF package which models a galaxy image as the sum of two light profiles: e.g., a de Vaucouleurs (bulge) component and an exponential (disc) component. GIM2D fits are available for over a million galaxies in SDSS DR7 \citep{sim10}, making it a useful tool for a large statistical study. GIM2D has been used to study the properties of bulge and disc components separately in galaxies from the Millennium Galaxy Catalogue \citep{all06,dri07,cam09}. \citet{sim02} have shown that GIM2D's bulge fraction $B/T$ and smoothness $S$ can be used to identify early type galaxies. \citet{im02} used these criteria to obtain a sample of field E/S0s in the DEEP Groth Strip Survey \citep{wei05,vog05}. In addition, \citet{mci04} used $S$ to study the presence of substructure in cluster disc galaxies. These two parameters, along with the SDSS-measured concentration $C$, are a natural starting point for our automated method. We will also find that the SDSS-measured axis ratio $b/a$ is useful in distinguishing the different morphologies.

As described below, we initially use emission line strengths and line ratios from SDSS spectra to weed out most of the objects that are obviously star-forming. After requiring that the targets be included in a number of external catalogs (to facilitate future studies), we isolate a sample of roughly one thousand red sequence galaxies. These are classified \textit{by eye} into three main groups based on the light profile of the galaxy at large radii: bulges, smooth discs, and unsmooth discs, the last group showing a variety of features. In addition to evaluating whether a galaxy is dominated by a bulge or disc component, we also judge whether it has any additional features, such as spiral arms or dust lanes, which would signal a cold disc.

The main goal of this paper is to compare these visual classifications with machine-derived structural parameters from SDSS and GIM2D. This allows us to develop a method for reproducing the visual classifications using automated parameters. With the benefit of our detailed visual inspections, we are able to take certain sets of machine parameters \textit{in combination} to isolate samples of bulge- and disc-dominated galaxies that agree with our visually classified samples. The resulting recipes can readily be applied to a much larger sample of early type galaxies from SDSS in order to classify them by their morphologies with high accuracy. Future use of this method may be possible with other samples with similar physical resolution scales, though some testing on these new samples will be required.

Finally, the sample of a thousand objects is used to derive some preliminary conclusions on the frequencies of bulges and discs as a function of magnitude, stellar mass, velocity dispersion, and radius. The results are verified using both the visual and automated classification schemes. 

The paper is organized as follows: \S\ref{selection} describes the sample of red sequence galaxies used in this study. These galaxies are grouped into bulges, smooth discs, and unsmooth discs using our visual classification scheme (\S\ref{visualscheme}). We then compare automated parameters -- SDSS photometric values (\S\ref{sdssparams}) and GIM2D parameters (\S\ref{gim2dparams}) -- with the visual classifications to develop an automated recipe for sorting galaxies into bulges, smooth discs, and unsmooth discs. \S\ref{prelimconc} summarizes some preliminary conclusions based on the visual classifications. Galaxies are next sorted into automated bulge and disc samples (\S\ref{autobulgedisc}) and the latter group is divided into smooth and unsmooth discs (\S\ref{smoothunsmooth}). In \S\ref{results} we present the bulge/disc frequencies as a function of absolute magnitude, stellar mass, velocity dispersion, and radius. We summarize the work in \S\ref{summary}. Some additional notes on the automated parameters are discussed in \S\ref{appendix}. The adopted cosmology is $\Omega_{\Lambda}=0.7$, $\Omega_{m}=0.3$, and $H_{0}=70$ km  s$^{-1}$ Mpc$^{-1}$. All absolute magnitudes and colors have been k-corrected to $z=0.0$ using Michael Blanton's \textit{k-correct} code (\citealp{bla03}, v3$\_$2).

\section{Data}\label{data}

\subsection{Sample selection}\label{selection}
The sample is extracted from the spectroscopic catalog of galaxies in Data Release 4 (DR4) of the SDSS \citep{ade06}. Because we are interested in studying quenched galaxies, we apply emission line criteria, based on equivalent width (EW) measurements from \citet{yan06}, to select galaxies whose spectra do not reveal obvious ongoing star formation. Such galaxies are defined as either lacking detectable emission (quiescent) or having LINER-like emission line ratios according to the criteria of \citet{gra07}. Quantitatively, galaxies lacking emission satisfy the conditions EW[O{\small II}] $<$ 3 \AA, EW(H$\alpha$) $<$ 0.7 \AA, while LINER-like galaxies satisfy the conditions EW[O{\small II}] $>$ 3 \AA, EW[O{\small II}] $> 5\times$ EW(H$\alpha$)$-7$. The latter criterion follows from the results of \citet{yan06}, who showed that galaxies with high [O{\small II}]/H$\alpha$ ratios (LINERS) are not actively forming stars. Galaxies with EW[O{\small II}] $>$ 3 \AA and low [O{\small II}]/H$\alpha$ ratios contain star formation and/or Seyfert activity, and we do not include them in our sample. 

We choose only galaxies with spectra having $S/N>20 \mbox{ \AA}^{-1}$ in order to have a sample with reliably measured equivalent widths. Of the remaining galaxies, $\sim90 $ per cent have specific star formation rates less than 0.1 M$_{\odot}$ yr$^{-1}$/10$^{11}$M$_{\odot}$ and are predominantly located on the red sequence or in the green valley, though there is a tail of bluer objects.\footnote{\footnotesize{Star formation rates are derived from H$\alpha$ emission line luminosities following Equation 2 of Kennicutt 1998.}} The bluest objects in this sample are removed by requiring that the sample satisfy the relation $g - r > -0.025r + 0.1$.\footnote{\footnotesize{During the initial sample selection, the SDSS magnitudes used were not corrected for Galactic extinction. Consequently, there are six galaxies that do not satisfy this color cut once the corrected magnitudes are used. The corrected magnitudes are used for all analysis in this paper, and this issue only comes up in the sample selection.}} There are known issues with SDSS photometry for large galaxies (e.g., \citealt{lau07}), which causes some extended blue discs to have red colors in the database. Because the color cut that we impose is generous, problems with the $g-r$ photometry will not affect the sample greatly.

We acknowledge that our sample selection is not perfect, and there will be some star forming galaxies that make it through our selection criteria. These interlopers are present mainly because SDSS spectra are obtained using $3''$ fiber apertures. Some of the galaxies in our sample may be forming stars in their outer regions but satisfy our emission line criteria because their non-starforming bulges dominate the spectral fiber aperture. 

Even with improved photometry from GIM2D, the $g-r$ color is not sensitive enough to identify these star-forming galaxies. A much better way to find star-forming contaminants is to pick out the bluest galaxies in $NUV-r$, using GALEX photometry. Though we could eliminate the contaminants from the sample presented here, we elect to keep these objects because adding a selection criterion using GALEX photometry would significantly limit the size of future samples of SDSS galaxies that could be examined with our automated classification scheme. Instead, we will show that most of these galaxies are unsmooth discs in our schema and can be identified and removed using the automated scheme. Throughout the analysis (see \S\ref{smoothunsmooth}, \S\ref{results}) we will return to this issue to confirm that this choice does not affect the results.

We would also like these data to be useful for future work exploring how morphology correlates with UV colors and environment. For the former condition, we limit the sample to galaxies with a near-UV detection in Data Release 2 (DR2) of the Galaxy Evolution Explorer (GALEX)\footnote{\footnotesize{http://galex.stsci.edu/GR2/}}. For the latter condition, we limit the sample to galaxies with measured overdensities in the NYU Value-Added Catalog \citep{bla05a}. All of these galaxies also have another measure of environment, the $1+\delta_{3}$ parameter of \citet{coo08}.

In numbers, the sample selection is as follows:

\begin{itemize}
\item \citet{yan06} emission line measurements (SDSS DR4): $\sim400,000$ galaxies
\item Falls in GALEX DR2 footprint: 29,755
\item NYU Value-Added Catalog Environments (SDSS DR2): 8,865
\item Median signal-to-noise ratio $> 20 \mbox{ \AA}^{-1}$: 3,331
\item Quiescent (915) or LINER (483): 1,398
\item $g-r$ color cut (894 + 447): 1,341
\item GALEX NUV detection (870 + 425): 1,295
\end{itemize}

Neither the $g-r$ color cut nor the NUV detection cut removes more than 5 per cent of the sample, so we do not believe that they have introduced any important selection effects. The surviving 1,295 galaxies fall in the redshift range $0.024 < z < 0.082$. The lower limit is set by the O{\small II} detection requirement: the line is located at $3727 \mbox{ \AA}$ while the SDSS spectra begin at $3800 \mbox{ \AA}$. The upper limit is set by the NYU Value-Added Catalog, which only has environment measured for galaxies with $z < 0.082$.

\subsection{Visual classification scheme}\label{visualscheme}

The visual classifications are based on images obtained from the SDSS Image List Tool,\footnote{\footnotesize{http://cas.sdss.org/astro/en/tools/chart/list.asp}} which generates multi-band color JPEG thumbnails of each galaxy. (See \citealt{lup04} and \citealt{nie04} for details on how these images are made.) The images were examined on a computer monitor (as opposed to printed pages) in order to see the full dynamic range.

\begin{figure*}
\includegraphics[width=1.0\textwidth]{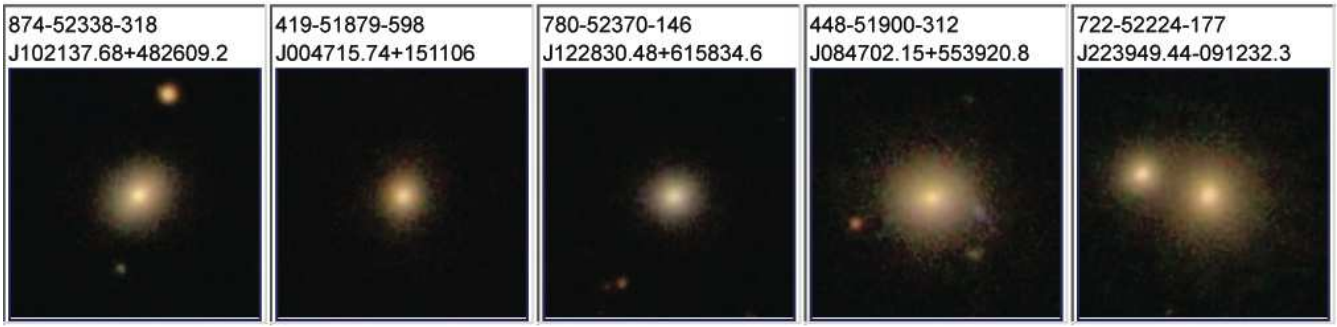}
\caption{Examples of visual bulges ($BC = 1$). These are objects that are highly centrally concentrated and have \textit{no sharp outer edge}. Each galaxy is labelled with its SDSS plate, modified Julian date, and fiber ID. The postage stamps are approximately 50 x 50 arcsecs. For visual classification, the images were examined on a computer screen in order to see the full dynamic range.}
\label{bdeq1}
\end{figure*}

\begin{figure*}
\includegraphics[width=1.0\textwidth]{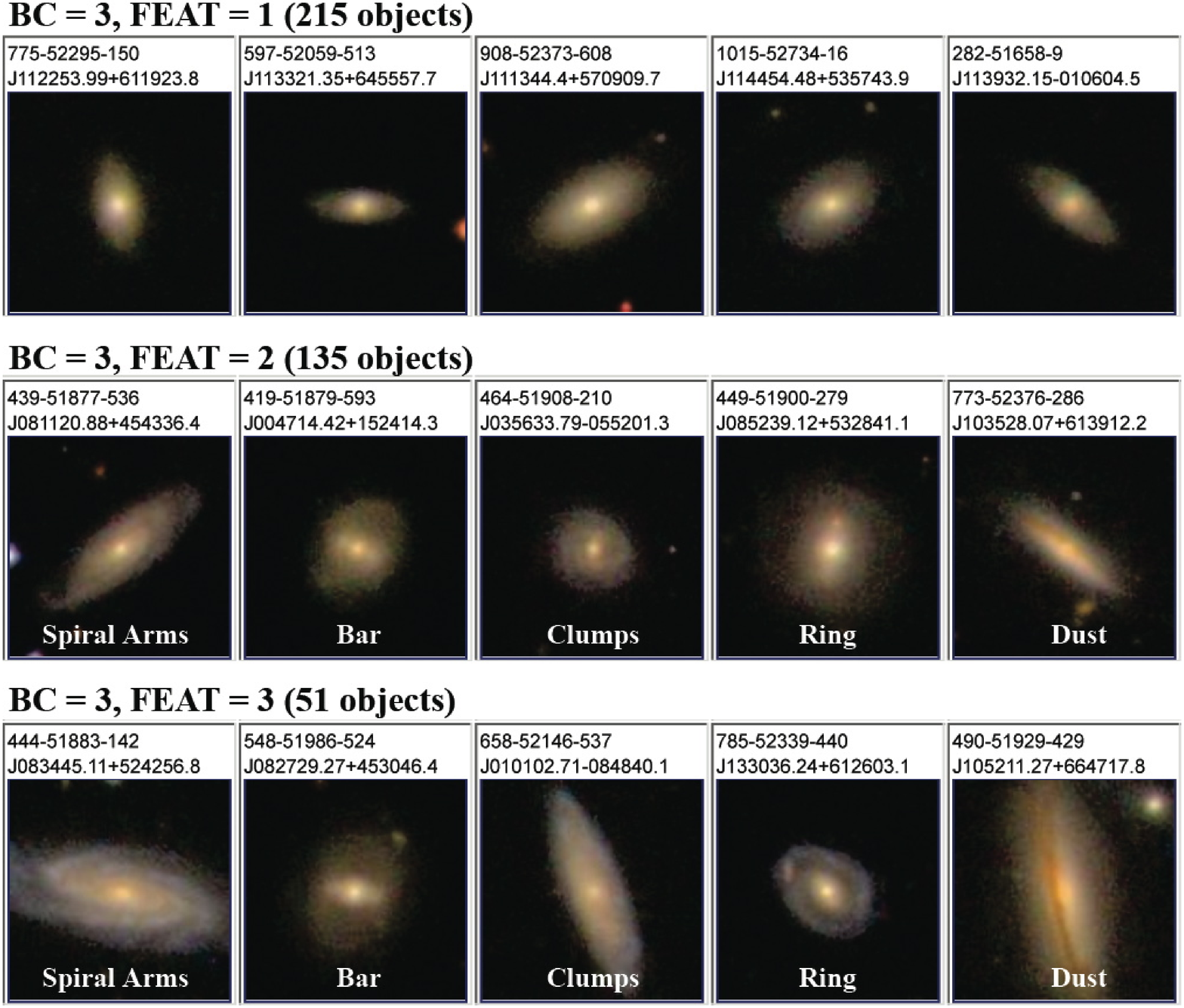}
\caption{Examples of visual discs ($BC = 3$). These are objects that have a \textit{sharp outer edge} where the light drops off rapidly, with a flatter light profile at intermediate radii just inside the outer boundary. Row 1: Smooth, featureless discs ($FEAT = 1$). Row 2: Unsmooth discs with weak features ($FEAT = 2$). Row 3: Unsmooth discs with strong features ($FEAT = 3$). Columns show the kinds of features noted in our visual classification scheme, from left to right: Spiral Arms, Bars, Clumps, Rings, and Dust. Some features, such as clumps and dust, show better on a computer monitor and are not reproduced well on paper.}
\label{bdeq3}
\end{figure*}

The first visual classification parameter sorts galaxies according to their radial light profiles; this is the bulge class, $BC$. Bulge-dominated galaxies ($BC = 1$) have bright centres with a very gradual fall-off in brightness at all radii. The outer regions are characterized by an extended envelope \textit{without a clear edge}. Examples are shown in Figure~\ref{bdeq1}. Disc-dominated galaxies ($BC=3$), on the other hand, have a sharp outer edge where the light drops off dramatically, as well as a flatter light profile at intermediate radii just inside the outer boundary. Examples are shown in Figure~\ref{bdeq3}. The stretch used for the thumbnails in the SDSS Image Tool is very tight near the sky level and shows these differences to good advantage. 

We explicitly acknowledge that our $BC = 1$ galaxies are not pure bulges, and our $BC = 3$ galaxies are not pure discs. For that reason \textquoteleft bulge-dominated' and \textquoteleft disc-dominated,' respectively, would be more appropriate terms. However, for brevity, we will use the terms \textquoteleft bulges' and \textquoteleft discs' in the rest of the paper.

\begin{figure*}
\includegraphics[width=1.0\textwidth]{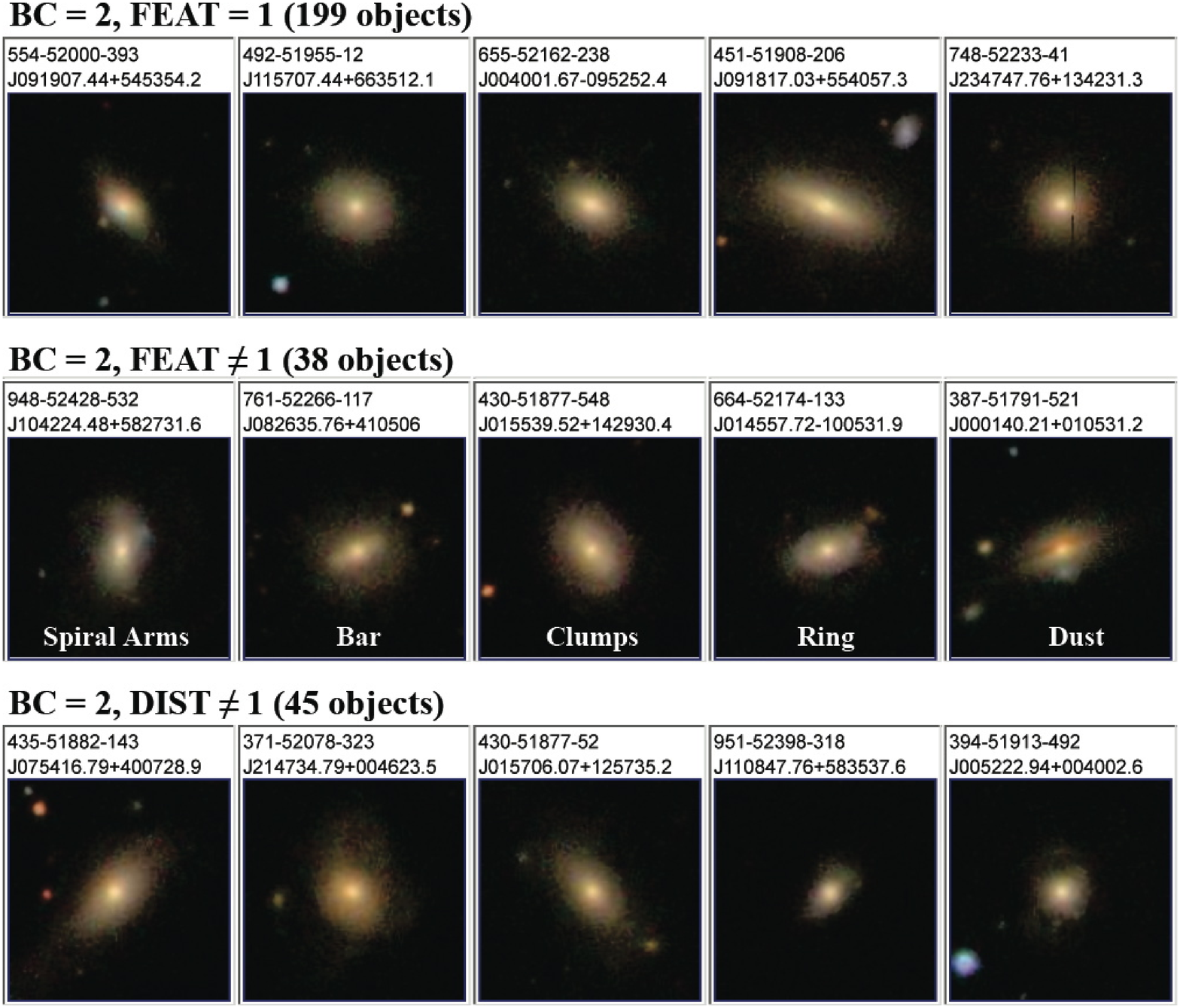}
\caption{Examples of visual intermediates ($BC = 2$). These galaxies are judged to be intermediate between the visual bulge and disc categories. Row 1: Smooth, featureless galaxies ($FEAT=1$) with light profiles that are not obviously bulgy or discy. Row 2: Same as Row 1, but with features ($FEAT\ne1$). All types of features found in our visual classification scheme are seen in intermediate galaxies; most of the features are weak. Row 3: Examples of problematic objects that have been classified as intermediate due to their disturbed morphologies, which make the edge profiles hard to assess.}
\label{bdeq2}
\end{figure*}

Objects that do not fall clearly into either the bulge or disc categories are classified as intermediate ($BC = 2$). Examples are shown in Figure~\ref{bdeq2}. Many have characteristics indicating the presence of both bulge-like and disc-like components. These include galaxies with a gradual fall-off of light in some azimuthal sectors but also a sharp boundary in others (these are often edge-on or disturbed). Others are objects that appear to have a disc embedded within a more extended classical bulge. Many such discs tend to be edge-on, which suggests that we tend to identify such objects preferentially when they are highly inclined. (See \S\ref{autoscheme} for more discussion.) We stress, however, that the identification of a disc depends purely on finding either sharp-edged or flattened (edge-on) features in the \textit{light} distribution, and thus the presence of a dust lane or other features related to the classification parameter $FEAT$ (described below) has nothing to do with the value of $BC$. 

The values for $BC$ were first determined by J. C. and then agreed upon by consensus among J. C., G. G., and S. F. Between the first independent classification and the group consensus there was remarkably little disagreement ($\sim3 $ per cent of the total sample had their $BC$ values changed). Objects for which there was disagreement were always on the borderline between the intermediate ($BC = 2$) category and the extremes ($BC = 1$ or 3). In no case was there disagreement between the two extremes. The number of objects moved in each direction when adopting final classifications was about equal. That is, the number of objects moved from bulge to intermediate ($BC=1$ to 2) was about equal to the number of objects moved from intermediate to bulge ($BC=2$ to 1).

When developing our visual scheme, a brief attempt was made to further distinguish between more bulge-like and more disc-like intermediates (which we coined \textquoteleft 2-' and \textquoteleft 2+', respectively), so as to make a 5-point scale in $BC$ instead of a 3-point scale. The resulting classifications yielded small samples of each new type, and the great bulk of intermediates remained in the middle bin ($BC = 2$). Because the new classifications affected such a small number of the galaxies, the 3-point scale was kept for simplicity. This will be revisited for the analysis in \S\ref{results}.

The second visual classification parameter sorts galaxies according to the presence of non-smooth \textquoteleft features': spiral arms, bars, clumps, rings, or dust. Most galaxies have no visible features ($FEAT = 1$). Other galaxies have features that we identify as weak ($FEAT = 2$) or strong ($FEAT = 3$). The latter are galaxies in which arms, bars, rings, and dust lanes can be traced over a considerable extent and/or individual clumps are highly visible. Examples of the different features and their varying strengths are shown in Figure~\ref{bdeq3}.

For each galaxy, the strength of each type of feature is noted (absent, weak, or strong), resulting in five different numerical values between 1 and 3. The $FEAT$ index is determined by taking the maximum of these five values. For example, a galaxy with strong spiral arms and a weak bar  is classified as $FEAT=3$. A galaxy with weak spiral arms and weak dust lanes is classified as $FEAT=2$. A galaxy has $FEAT=1$ only if it exhibits none of the five features.

The $FEAT$ index encompasses two types of features. One type is classically identified with star formation -- spiral arms, clumps, and dust absorption. The second type -- bars and rings -- may exist in otherwise smooth discs without obvious star formation. A noteworthy finding in this work is that, with the rare exception of dust, none of the above feature elements are found to any significant degree in \textit{bulge-dominated} (i.e., $BC = 1$) galaxies (cf. Figures~\ref{bdeq1} and~\ref{bdeq3}). In other words, as expected, the presence of any of these features, of whatever kind, is an excellent predictor of a \textit{disc-dominated} (i.e., $BC = 3$) galaxy, and the close correlation is independent confirmation of our purely light-based bulge-disc classifications  ($BC$). 

Since our focus in this paper is on the relative numbers of bulges and discs, we have simply lumped together all types of features to define the $FEAT$ index, though the exact nature(s) of the features detected (whether bar, spiral, etc.) is retained in the database. These features are also difficult to distinguish using automated parameters, so we elect to use only one visual index. While not all of the features are associated with star formation, most galaxies with $FEAT\neq1$ (78 per cent) have at least one starforming feature. Thus, we will use $FEAT$ as an indicator of galaxies that are not truly quenched. 

We also separately identify any galaxies that are disturbed or interacting (some are shown in Figure~\ref{bdeq2}) with indices $DIST$ and $INT$. These make up a small portion of the sample (see \S\ref{prelimconc}) and are not used in the development of the automated method (\S\ref{autoscheme}). As with the other visual parameters, the values are determined by eyeball classification on a discreet scale from 1 to 3, with $DIST (INT) = 1$ being undisturbed (not interacting) and $DIST (INT) = 3$ being strongly disturbed (obviously interacting). Whereas $FEAT$ measures the strengths of very specific types of morphological features, $DIST$ and $INT$ are completely separate from $FEAT$ and measure the peculiarities in a galaxy's appearance or immediate environment. 

A major limitation of the visual scheme is the difficulty of classifying small galaxies. These galaxies appear generally compact, and little or no structure is visible. Because of their small size, the shape of the outer light profile is difficult to estimate visually, and no substructures can be resolved. If the radius encompassing 90 per cent of the Petrosian flux ($R_{90}$) is used as a proxy for size, these galaxies have small $R_{90}$ compared to the rest of the sample, typically with values log $R_{90}\le1.15$ ($R_{90}\sim 14''$). This criterion is able to identify 75 per cent of the galaxies that were visually determined to be too small to classify reliably. We make the size cut using an automated parameter to allow for future application to larger samples. After removing galaxies with log $R_{90}\le1.15$, we are left with 998 galaxies in the sample.

\subsection{SDSS photometry}\label{sdssparams}
Of these galaxies, 997 had full SDSS photometric fits, which are available as part of SDSS DR4. For each galaxy, we obtained the model magnitudes in the SDSS $g$- and $r$-bands, as well as the axis ratio in the $r$-band ($b/a$). The model magnitudes are calculated using either a de Vaucouleurs or exponential profile, with the choice of profile being determined by likelihoods calculated from $\chi^{2}$ fits. The axis ratio is always determined from the de Vaucouleurs fit. More detail on how these quantities are measured can be found in the description of the SDSS Data Release 2 \citep{aba04}. We also utilize the radii containing 50 per cent and 90 per cent of the Petrosian flux in the $r$-band ($R_{50}$ and $R_{90}$; \citealp{bla01}). The ratio $R_{90}/R_{50}$ is the concentration $C$.

\subsection{GIM2D parameters}\label{gim2dparams}
The galaxy images were run through GIM2D, which fits the light of a galaxy as the sum of a de Vaucouleurs (bulge) profile and an exponential (disc) profile \citep{sim02}. The best-fit bulge+disc decomposition is found through a $\chi^{2}$-minimization of 12 free parameters, as described by \citet{sim02}. We use two of these parameters in our analysis: the bulge light fraction ($B/T$) and the disc inclination ($i$). The bulge fraction is given by the flux from the bulge component divided by the total flux from both the bulge and disc components. A pure disc has $B/T=0$. The disc inclination is the angle between the vertical axis of the fitted disc and the line of sight, in degrees. A face-on disc has $i = 0$. 

In addition, GIM2D also computes image indices from the residual image (galaxy image minus model fit). Two of these indices, the total residual ($R_{T}$) and the asymmetric residual ($R_{A}$), measure the galaxy's deviation from the smooth model light profile. Quantitatively, these are defined in Equation 11 by \citet{sim02}:

\begin{equation}
R_{T}=\frac{\Sigma(1/2)|R_{ij}+R_{ij}^{180}|}{\Sigma I_{ij}}-\frac{\Sigma(1/2)|B_{ij}+B_{ij}^{180}|}{\Sigma I_{ij}}
\end{equation}

and

\begin{equation}
R_{A}=\frac{\Sigma(1/2)|R_{ij}-R_{ij}^{180}|}{\Sigma I_{ij}}-\frac{\Sigma(1/2)|B_{ij}-B_{ij}^{180}|}{\Sigma I_{ij}},
\end{equation}

\noindent where $R_{ij}$ is the pixel value of an object pixel in the residual image, $I_{ij}$ is the corresponding pixel value in the science image, and $B_{ij}$ is the pixel value of a random background pixel in the science image. $R_{ij}^{180}$ and $B_{ij}^{180}$ are the corresponding galaxy and background pixel values in the residual image after a 180 degree rotation about the centre of the galaxy. The position of the centre is specified by two of the 12 free parameters that are fit in GIM2D. We were able to obtain GIM2D parameters for 986 galaxies. The GIM2D fits used here were derived by simultaneous fitting of $g$- and $r$-band images, SDSS deblending, and sky levels determined by GIM2D; the details are described in \citet{sim10}.

In our analysis, we will be using the sum of the residuals $R_{T}$ and $R_{A}$, called the smoothness parameter $s2$. The \textquoteleft2' indicates that $R_{T}$ and $R_{A}$ are calculated within two half-light radii of the galaxy. Actually, a better name for $s2$ would be the \textquoteleft unsmoothness' parameter, as high $s2$ indicates a clumpy light distribution (large $R_{T}$), the presence of asymmetric features (large $R_{A}$), or both. 

We remove two galaxies with suspicious values of $s2$. Galaxies with a bright star or galaxy companion near their centres fall in the tails of the $s2$ distribution. Some of the galaxies with the lowest values of $s2$ have large companion galaxies superimposed. Two objects have $s2<0.0$, and both are overwhelmed by the light of the nearby, larger galaxy. This makes the background subtraction difficult, which causes the value of $s2$ to be negative. These fits appear to be unreliable, so we omit all targets with $s2 < 0.0$ from the sample.

Some of the galaxies with the highest values of $s2$ have bright stars visible in the field, and their high $s2$ values may be an artefact of the superimposed extra object. These are difficult to distinguish from objects with genuinely unsmooth features with a simple cut in $s2$ (unlike the case of objects with negative $s2$), so we leave them in the sample.

The final sample consists of 984 galaxies that are large enough to classify visually and have valid automated parameters in the redshift range $0.02 < z < 0.08$. Because the galaxies cover a relatively small redshift range, errors from the K-correction should be small. Table~\ref{datatable} presents visual classifications, SDSS identifiers and photometry, GALEX photometry, and GIM2D parameters for the sample. A full electronic version of the table is available.

\begin{table*}
\caption{Visual Classifications and Automated Parameters for 984 Galaxies.}
\begin{tabular}{lcccccccccccccccccccccccccccc}
\hline
(1) & (2) & (3) & (4) & (5) & (6) & (7) & (8) & (9) & (10) & (11) & (12) & (13) & (14) & (15) & (16) & (17) & (18) & (19) & (20) & (21) & (22) & (23) & (24) & (25) & (26) & (27) & (28) & (29)\\
\hline
718 &  52206  & 182 &  587726877271457954  & 332.419 &  -9.430 & -19.721&  -21.562 & -22.295  & -22.720 & -23.000 & -16.616 & -236.029 &   7.057 &  23.240 &  0.886  &  0.838  &  0.028 &  32.605 & 1&  1 & 1 & 1 & 1 & 1 & 1 & 1 &  1 & 1\\
558 & 52317 & 410 & 588007003632435285 & 148.870 & 58.890 & -19.367 & -21.158 & -21.890 & -22.262 & -22.545 & -16.649 &  -236.033 & 5.424 & 17.746 & 0.536 & 0.776 & 0.049 & 59.659 & 2 & 1 & 1 & 1 & 1 & 1 & 1 & 2 & 1 & 2\\
 381 & 51811 &  74 & 587731186198708399 & 348.134 & -0.110 & -18.065 & -19.810 & -20.538 & -20.881 & -21.151 & -15.160 & -13.072 &  5.642 & 17.604 &  0.489 &  0.766 &  0.101 & 69.418 & 3 & 1 & 1 & 1 & 1 & 1 & 1 & 1 & 1 & 3\\
 438 & 51884 & 284 & 587725775070036188 & 120.059 & 46.043 & -19.258 & -20.967 & -21.851 & -22.086 & -22.455 & -17.694 & -16.387 & 10.798 & 22.581 &  0.468 &  0.094 &  0.114 & 61.389 & 3 & 2 & 1 & 1 & 2 & 1 & 1 & 1 & 1 & 3\\
 771 & 52370 & 341 & 587725475491610665 & 150.930 & 61.739 & -18.730 & -20.646 & -21.415 & -21.814 & -22.110 & -15.574 & -234.919 &  5.861 & 17.899 &  0.595 &  0.890 &  0.047 & 41.486 & 1 & 1 & 1 & 1 & 1 & 1 & 1 & 1 & 1 & 2\\
\hline
\label{datatable}
\end{tabular}
Full version available online. Columns: (1-11) SDSS Plate, MJD, Fiber ID, ObjID, RA, Dec, K-corrected $ugriz$ ; (12-13) GALEX K-corrected $NUV, FUV$; (14-16) SDSS Petrosian R$_{50}$, Petrosian R$_{90}$, Axis Ratio $b/a$; (17-19) GIM2D bulge fraction $B/T$, smoothness $s2$, inclination $i$; (20-28) Visual Classification $BC, FEAT, ARMS, BAR, CLUMPS, RING, DUST, DIST, INT$; (29) Automated Classification: 1-Bulge, 2-Intermediate, 3-Disc
\end{table*}

\subsection{Preliminary conclusions based on visual classifications}\label{prelimconc}
In this section we present some findings based on the visual classifications of the final sample of 984 galaxies. First, a large fraction of these early type galaxies are found to have strong discs. Our sample consists of 346 bulges (35 per cent, $BC=1$), 237 intermediates (24 per cent, $BC=2$), and 401 discs (41 per cent, $BC=3$). Thus, our analysis confirms what had been known before, that S0s and Sas (i.e., discs) comprise a very significant fraction of red sequence galaxies.

\begin{figure*}
\includegraphics[width=1.0\textwidth]{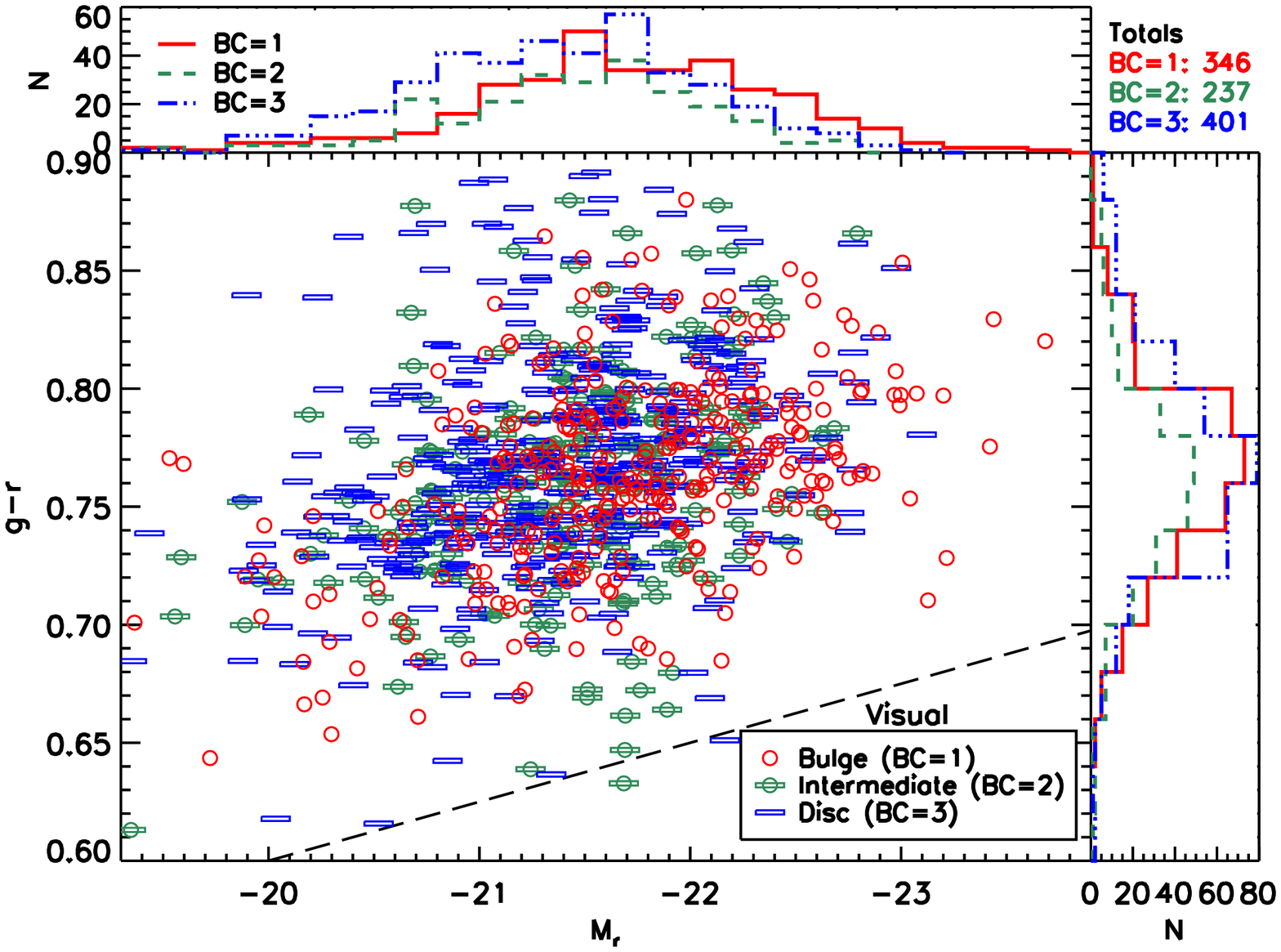}
\caption{Color magnitude diagram of the final sample of 984 galaxies that are large enough to classify visually and have valid automated parameters. Absolute magnitudes are SDSS model fluxes k-corrected to $z=0.0$ using Michael Blanton's \textit{k-correct} code (\citealp{bla03}, v3$\_$2). The symbols indicate the visual classification $BC$: bulges ($BC = 1$, red circles), discs ($BC = 3$, blue bars), and intermediates ($BC = 2$, green circles+bars). The dashed line indicates the color cut that was described in \S\ref{selection}. Using only the emission line criteria described in the text to remove star forming galaxies yields a sample of predominantly red and green galaxies. The top and right panels show histograms of each visual type in magnitude and color, respectively. The visual bulges (red) occupy a large range in magnitude but a narrower range in color compared to the visual discs (blue dash-dotted) and intermediates (green dashed).}
\label{cmd}
\end{figure*}

Figure~\ref{cmd} shows a color-magnitude diagram of the sample, with colors and symbols denoting bulge class $BC$. The panels at the top and right show the magnitude and color distributions, respectively, of bulges, intermediates, and discs. While the colors of all three classes peak at about the same value, the bulges have the narrowest distribution, and most of the outliers are intermediates or discs.

As expected, most of the brightest objects are giant ellipticals, and most of the faintest objects are discs. There are, however, also many bright discs. At magnitudes brighter than $M_r=-22$, there are 121 bulges ($BC=1$), 41 intermediates ($BC=2$), and 69 discs ($BC=3$). The objects in this last group are truly discs: 48 of the 69 discs brighter than $M_r = -22$ have features ($FEAT\ne1$), which would likely cause them to be classified as Hubble Types Sa or Sb. We will be returning to these galaxies in \S\ref{autobulgedisc}.

Second, 229 (23.3 per cent) galaxies show features ($FEAT\neq1$). The frequency of each feature type is listed in Table~\ref{featnum}, which shows the number of galaxies with each $BC$ with various features at different strengths. Very few bulges ($BC = 1$) have features, and those that do have dust. In contrast, $\sim80 $ per cent of face-on $BC=3$ galaxies show features, nearly all of which are associated with gas and/or dynamically cold stellar populations, both of which are well-known disc tracers. The evidence from $FEAT$ is therefore a strong independent validation that our main disc criterion based on sharp outer brightness profile fall-off is indeed highly correlated with the presence of a dynamically cold, rotating stellar population.

\begin{table*}
\begin{minipage}{110mm}
\caption{Frequency of Features for 984 Galaxies.}
\begin{tabular}{lccc|ccc|ccc|ccc}
\hline
 & \multicolumn{3}{c}{$BC=1$} &  \multicolumn{3}{c}{$BC=2$} &  \multicolumn{3}{c}{$BC=3$} & \multicolumn{3}{c}{cos $i > 0.8$}\\
Feature & 1 & 2 & 3 & 1 & 2 & 3 & 1 & 2 & 3 & 1 & 2 & 3\\
\hline
$ARMS$ & 346  & 0 & 0 & 236 & 1 & 0 & 341 & 47 & 13 & 59 & 14 & 7\\
$BAR$ & 346 & 0 & 0 & 225 & 11 & 1 & 316 & 70 & 15 & 52 & 20 & 8\\
$CLUMPS$ & 346 & 0 & 0 & 222 & 14 & 1 & 302 & 89 & 10 & 39 & 39 & 2\\
$RING$ & 346 & 0 & 0 & 233 & 4 & 0 & 329 & 57 & 15 & 50 & 22 & 8\\
$DUST$ & 341 & 3 & 2 & 226 & 8 & 3 & 376 & 19 & 6 & 79 & 1 & 0\\
\hline
$FEAT$ & 341 & 3 & 2 & 199 & 33 & 5 & 215 & 135 & 51 & 19 & 41 & 20\\
\hline
\hline
$DIST$ & 328 & 15 & 3 & 192 & 33 & 12 & 365 & 28 & 8 & 69 & 7 & 4\\
$INT$ & 344 & 2 & 0 & 231 & 6 & 0 & 396 & 4 & 1 & 78 & 1 & 1\\
\hline
\label{featnum}
\end{tabular}
The upper part of the table shows the numbers of objects with the five kinds of features that make up the $FEAT$ classification, which collects together all types of features that might signal a disc. For each visual morphological type, the number of galaxies with each feature strength is shown. The last column (cos $i>0.8$) lists the frequencies for face-on discs only. Note that visual bulges ($BC=1$) show no features except dust lanes. The lower part of the table shows the number of objects with each value of $DIST$ and $INT$. About 10 per cent of the sample is disturbed or interacting with another galaxy.
\end{minipage}
\end{table*}

Finally, only 99 galaxies ($\sim10 $ per cent) appear to have disturbed morphologies, and only 13 ($\sim1.3 $ per cent) appear to be interacting with nearby neighbors. Nearly half of these were classified as intermediates ($BC = 2$), as their disturbed appearances often made it difficult to identify them as having clear bulge or disc morphologies. However, we can already see that the SDSS thumbnails are not the best material to judge disturbances, and evidence of more subtle peculiarities is often difficult to see by eye. Future work on these objects may be done using GIM2D's residual images, which are a much more powerful tool for this kind of analysis. Furthermore, the limiting surface brightness of the SDSS is too bright to see the fainter tidal features such as those observed by \citet{van05}, whose much deeper imaging showed that 53 per cent of their red galaxy sample exhibited signatures of tidal interactions.

\section{Automated Classification Scheme}\label{autoscheme}

We have identified three populations -- bulges, smooth discs, and unsmooth discs -- in our sample of 984 galaxies using a visual classification system. In this section, we show that we are able to reproduce the visual classifications using a set of automated parameters derived from the models fit by GIM2D together with photometric measurements available from the data products of the SDSS. This automated classification method will allow us to assign morphologies to large numbers of SDSS galaxies in future work without having to visually inspect each one. We will refer to the resulting morphological samples as the \textit{automated bulge}, \textit{automated intermediate}, and \textit{automated disc} samples.

We initially assume that the visual classifications are the \textquoteleft true' classifications and determine what combinations of automated parameters are most effective in reproducing the visual morphological types. We divide the automated parameter space and assign an automated classification to each region based on which visual classification is most common in that region. With the automated classification scheme in place, we examine more closely the galaxies where the visual and automated classifications are discrepant. These examples shed light on the strengths and weaknesses of the visual classifications as well as on the accuracy of the machine-generated measurements made by GIM2D and SDSS. 

In addition, some instances in which the automated parameters are less reliable are discussed in \S\ref{appendix}. Briefly, $B/T$ is underestimated for luminous bulge-dominated galaxies that are more concentrated than a profile with S\'ersic index $n=4$ and $C$ is least reliable for the smallest and/or most elongated galaxies. Because our automated scheme uses several parameters in conjunction, the results presented in \S\ref{autobulgedisc} and \S\ref{smoothunsmooth} are unlikely to be strongly influenced by these findings.

\subsection{Bulges vs. discs}\label{autobulgedisc}

\subsubsection{Method}
\begin{figure}
\includegraphics[width=0.5\textwidth]{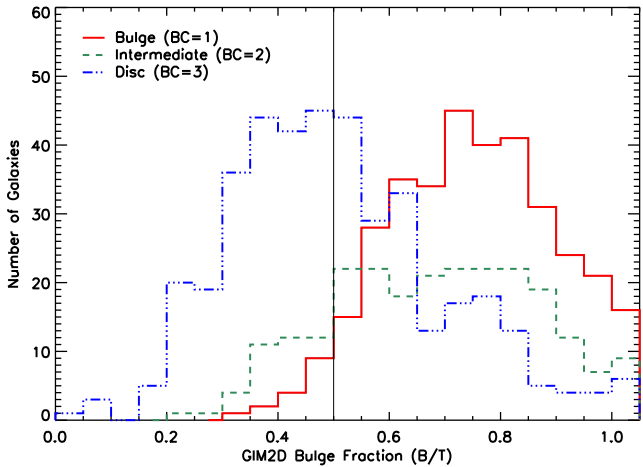}
\caption{Histogram comparing the GIM2D bulge fraction ($B/T$) and the visual classifications $BC$: bulges ($BC = 1$, red), discs ($BC = 3$, blue dash-dotted), and intermediates ($BC = 2$, green dashed). Most galaxies with $B/T\leq0.5$ are visual discs, but only about half of the galaxies with $B/T>0.5$ are visual bulges. The presence of a large number of visual discs and intermediates in the high-$B/T$ sample indicates that $B/T$ alone is not able to reproduce all of the visual classifications. The numerical results are tabulated in Table~\ref{bulgedisc}.}
\label{bfhist}
\end{figure}

Figure~\ref{bfhist} shows the correlation between the visual bulge class index ($BC$) and the GIM2D bulge fraction ($B/T$). As expected, visual bulges ($BC=1$, red) typically have higher $B/T$ than visual discs ($BC=3$, blue dash-dotted). We divide the sample into two populations, one with high bulge fraction ($B/T > 0.5$), which includes most of the visual bulges, and one with low bulge fraction ($B/T\leq0.5$), which includes most of the visual discs. We consciously set the $B/T$ boundary a bit low (at 0.5, rather than 0.55 or 0.6) to keep most of the visual bulges together in the same sample. Interloper discs and intermediates ($BC=2$, green dashed) will be removed from the high $B/T$ sample later using other automated parameters.

The above cut yields good general agreement between the visual classification $BC$ and the GIM2D parameter $B/T$, which indicates that both methods are able to distinguish between different light profiles with reasonable reliability. A comparison of the bulge and disc galaxies in Figures~\ref{bdeq1} and ~\ref{bdeq3} clearly shows the contrast between the two types. The bulges ($BC=1$, Figure~\ref{bdeq1}) appear more extended and have no clear outer boundary, while the discs ($BC=3$, Figure~\ref{bdeq3} Row 1) have a visible edge. These two groups of objects were identified initially by eye but have different values of $B/T$, which indicates that we are able to make the distinction between bulges and discs both visually and automatically. 

The number of each visual type falling on both sides of the $B/T=0.5$ cut are listed in Table~\ref{bulgedisc}. The completeness is the percentage of all visual bulges (discs) that are recovered by the $B/T$ criterion, while the purity is the percentage of all galaxies in the high-$B/T$ (low-$B/T$) sample that are truly visual bulges (discs). Based on these results, if given only $B/T$, we can reasonably say that galaxies with $B/T\le0.5$ are most likely visual discs; the purity of the low-$B/T$ sample is 79 per cent. Galaxies with $B/T>0.5$, however, cannot all be assumed to be visual bulges; half of high-$B/T$ galaxies are determined visually to have moderate ($BC=2$) or strong ($BC=3$) discs.

\begin{table}
\caption{Visual Bulge Class vs. GIM2D Bulge Fraction.}
\begin{tabular}{lccc}
\hline
Visual & \multicolumn{2}{c}{GIM2D} & Completeness\\ 
  & $B/T > 0.5$    & $B/T \leq 0.5$    &\\
\hline
Bulge ($BC = 1$)   &  \textbf{330}  &   16   &  95.4\\
Int. ($BC = 2$)   &  196    & 41   &  ---\\
Disc ($BC = 3$)   &  186   &    \textbf{215}   &  53.6\\
\hline
Purity   &  46.3 &   79.0\\
\hline
\label{bulgedisc}
\end{tabular}
Comparison of visual bulge class $BC$ and bulge fraction $B/T$. Bold values indicate instances where the methods agree. The completeness is the percentage of all visual bulges (discs) that are recovered by the $B/T$ criterion. The purity is the percentage of all galaxies in the high-$B/T$ (low-$B/T$) sample that are truly visual bulges (discs). The purity of the high $B/T$ sample is relatively low (46.3 per cent), which indicates that $B/T$ alone cannot reproduce the visual classifications.
\end{table}

\begin{figure*}
\includegraphics[width=1.0\textwidth]{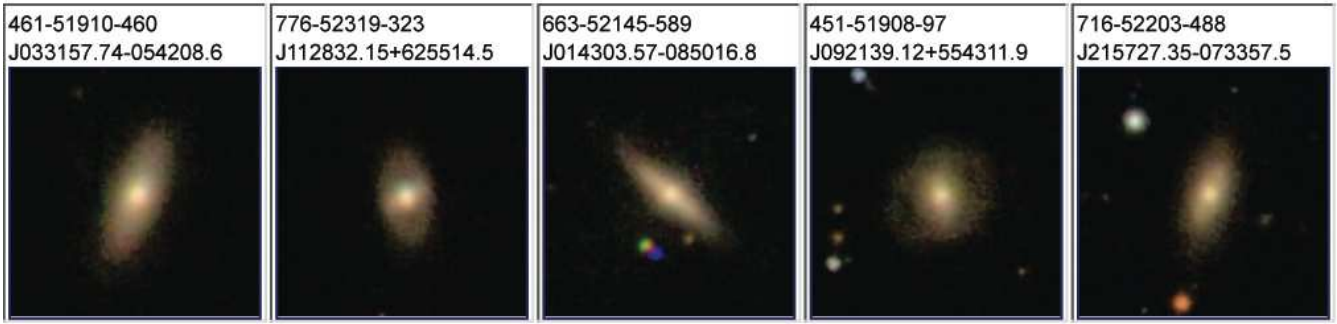}
\caption{Examples of visual discs ($BC=3$) that are bulge-dominated according to GIM2D ($B/T > 0.6$). All of these galaxies satisfy the criterion of having a clear edge, as described in the text. Visual inspection of these galaxies convinces us that high-$B/T$ galaxies are contaminated with a significant fraction of discs and thus that $B/T$ alone is not a reliable way to distinguish between bulges and discs.}
\label{bdeq3bfgt0.6}
\end{figure*}

To understand this disagreement we examine the visual discs with bulge fractions well above the $B/T=0.5$ boundary ($BC=3, B/T>0.6$); these make up 28 per cent of the visual disc sample. Examples are shown in Figure~\ref{bdeq3bfgt0.6}. Possible explanations for the disagreement are discussed below (\S\ref{disagreements}), but the point for now is that the great majority of these cases are indeed really disc-dominated systems and $B/T$ alone is insufficient to reproduce our visual classifications.

\begin{figure}
\includegraphics[width=0.5\textwidth]{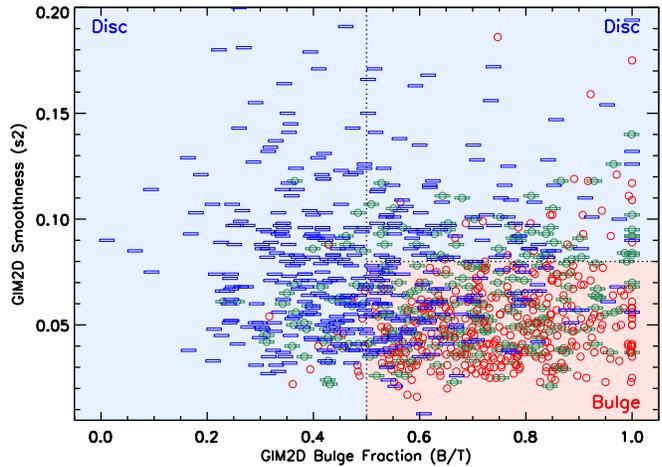}
\caption{GIM2D Bulge Fraction $B/T$ vs. GIM2D Smoothness $s2$. Symbols indicate visual classifications as defined in Figure~\ref{cmd}. Shaded regions indicate the adopted automated classifications. Of the high-$B/T$ galaxies, most visual bulges (red circles) are smooth ($s2 \le 0.08$), while visual discs and intermediates are more likely to show clumpiness or asymmetry ($s2 > 0.08$). We assign the high-$s2$ galaxies to the automated disc sample (blue shading). However, many visual discs and intermediates still remain in the low-$s2$ region containing candidates for the automated bulge sample (pink shading) which leads us to explore the additional cuts in Figures~\ref{bf_abr} and~\ref{bf_conc}.}
\label{bf_s2}
\end{figure}

Three other parameters allow us to distinguish the different morphologies within the high-$B/T$ sample: the GIM2D smoothness parameter $s2$, the SDSS axis ratio $b/a$, and the SDSS concentration index $C$. We apply cuts in these parameters successively using an algorithm that starts with $s2$, proceeds to $b/a$, and ends with $C$ (the order is discussed at the end of this section). We begin with Figure~\ref{bf_s2}, which shows $B/T$ plotted against $s2$ for the 984 galaxies in our sample. As in Figure~\ref{cmd} the symbols represent the three types of visual classification ($BC$): bulge (red circles), disc (blue bars), and intermediate (green circles+bars). The shaded regions represent the adopted automated classification boundaries, which will be described in more detail below.

All of the galaxies with $B/T\le0.5$ are classed as automated discs (shaded in blue), because this cut gives a relatively pure sample of discs (79 per cent, see Table 2). Figure~\ref{bf_s2} shows again that this region is clearly dominated by visual discs. We now turn our attention to the high-$B/T$ galaxies. We would like to use $s2$ to separate the bulges and discs with $B/T>0.5$.

For a given parameter (e.g., $s2$), the boundary between different types is chosen by eyeballing the value at which the local densities of visual bulges and discs are roughly equal. In general, this can be done by making a cut which places the tail of the visual bulge distribution into the automated disc sample, so that most visual bulges (with the exception of some outliers) are included in the automated bulge sample. A cut at $s2=0.08$ successfully distinguishes many of the visual intermediates and discs from the bulges in the high-$B/T$ sample, while also retaining most of the visual bulges in the automated bulge sample. We therefore designate the region $B/T>0.5, s2\leq0.08$ as containing \textit{candidates} for the \textit{automated bulge} sample (shaded in pink). These are as yet just candidates; their bulge status will be re-evaluated based on the values of the other automated parameters.

The next step is to determine what automated classification to assign to galaxies in the region $B/T> 0.5, s2>0.08$. Because most of the objects in this region are visual discs (90 of 169), we \textit{permanently} assign all galaxies in this region to the \textit{automated disc} sample (shaded in blue). The 79 discrepant objects include 48 visual intermediates and 31 visual bulges. By visual inspection, almost all of the latter have high smoothness parameters because of dust, disturbed shapes, nearby companions, or bright stars in the field. Thus, in assigning all of these objects to the automated disc sample, we make an error of at most one class for the great majority of objects and misclassify at most a handful of problematic bulges that would be difficult to deal with in any case.

\begin{figure}
\includegraphics[width=0.5\textwidth]{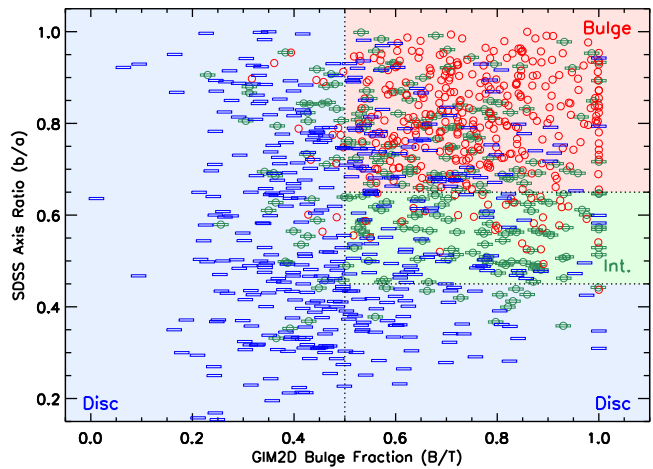}
\caption{GIM2D Bulge Fraction $B/T$ vs. SDSS Axis Ratio $b/a$. Symbols indicate visual classifications as defined in Figure~\ref{cmd}. Shaded regions indicate the adopted automated classifications. Of the high-$B/T$ galaxies, most visual bulges are round ($b/a > 0.65$) while visual discs and intermediates are more elongated ($b/a \le 0.65$). We assign the lowest-$b/a$ ($b/a\le 0.45$) galaxies to the automated disc sample (blue shading) and the intermediate-$b/a$ ($0.45\ge b/a <0.65$) galaxies to the automated intermediate sample (green shading). The possibility of a bias in the visual classifications due to inclination effects is discussed in the text (\S\ref{disagreements}).}
\label{bf_abr}
\end{figure}

Proceeding to the next step, Figure~\ref{bf_abr} shows $B/T$ plotted against $b/a$ for the 984 galaxies in our sample. Most visual bulges have high axis ratios (i.e., they are round), so objects in the region $B/T > 0.5, b/a > 0.65$ are designated as candidates for the automated bulge sample (shaded in pink). Based on the numbers of each visual type present, the regions $B/T>0.5, 0.45<b/a\leq0.65$ and $B/T>0.5, b/a\leq0.45$ are permanently assigned to the automated intermediate (shaded in green) and automated disc (shaded in blue) samples, respectively. More elongated galaxies are classed as visual intermediates or discs, consistent with traditional Hubble Types, in which elliptical galaxies cannot intrinsically have $b/a < 0.33$ (the most elongated ellipticals are E7). The possibility of a bias in our visual classifications due to inclination effects is discussed below (\S\ref{disagreements}). For now, we assume that our visual classifications are representative of the \textquoteleft true' morphologies.

\begin{figure}
\includegraphics[width=0.5\textwidth]{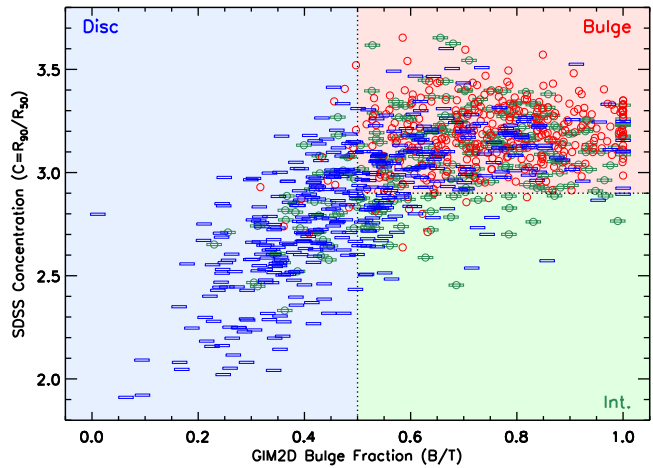}
\caption{GIM2D Bulge Fraction $B/T$ vs. SDSS Concentration $C$. Symbols indicate visual classifications as defined in Figure~\ref{cmd}. Shaded regions indicate the adopted automated classifications. Of the high-$B/T$ galaxies, most visual bulges are highly concentrated ($C > 2.9$). Most of the heretofore unclassified galaxies in the region $B/T > 0.5$, $C \le 2.9$ are visual intermediates, so we assign them to the automated intermediate sample (green shading). The appendix has further discussions of the high-$C$, small, and elongated galaxies in this figure.}
\label{bf_conc}
\end{figure}

Finally, Figure~\ref{bf_conc} shows $B/T$ plotted against $C$ for the 984 galaxies in our sample. The distribution of visual bulges (red circles) is consistent with the results of \citet{shi01} and \citet{str01}, who found that $C$ is closely correlated with a galaxy's morphology. Most visual bulges are highly concentrated, so objects in the region $B/T >0.5, C > 2.9$ are designated as candidate automated bulges (shaded in pink). Of the 87 objects in the region $B/T > 0.5, C\leq2.9$, 39 have already been classified as automated discs based on their values of $s2$ or $b/a$, and another 16 have been classified as automated intermediates based on their values of $b/a$. Of the remaining 32, 9 are visual bulges, 13 are visual intermediates, and 10 are visual discs. Because most of the unclassified objects in this region are visual intermediates, we assign them permanently to the \textit{automated intermediate} sample (shaded in green). Five of these galaxies were noted during visual classification as being small galaxies that were not eliminated from the sample using the log $R_{90}\leq1.15$ size cut described in \S\ref{visualscheme}. Additional discussion of the high-$C$, small, and elongated galaxies in this figure is given in the appendix.

To summarize, in Figures~\ref{bf_s2},~\ref{bf_abr}, and~\ref{bf_conc}, our visual bulges are generally round (high $b/a$), smooth (low $s2$), and centrally concentrated (high $C$). We therefore can say that a galaxy which occupies \textit{all three} of the pink shaded regions in Figures~\ref{bf_s2},~\ref{bf_abr}, and~\ref{bf_conc} is likely to be a visual bulge and should therefore be put in the automated bulge sample. 

Numerically, the \textit{automated bulges} satisfy the following criteria: $B/T > 0.5$, $s2 \leq 0.08$, $b/a > 0.65$, and $C > 2.9$. Similarly, objects are classified as \textit{automated discs} if they satisfy \textit{one} of the following criteria: $B/T\leq0.5$ \textit{or} $s2 > 0.08$ \textit{or} $b/a\leq0.45$. Objects are classified as \textit{automated intermediates} if they satisfy $B/T > 0.5$, $s2\le0.08$, $b/a>0.45$ and ($C\leq2.9$ \textit{or} $b/a\leq0.65$). In words:

\begin{enumerate}
\item{To be an automated bulge, a galaxy must have a high bulge fraction \textit{and} be smooth \textit{and} roundish.}
\item{To be an automated disc, a galaxy \textit{either} has to have low bulge fraction \textit{or} be lumpy \textit{or} be elongated.}
\item{Automated intermediates are all other cases.}
\end{enumerate}

\begin{figure}
\includegraphics[width=0.5\textwidth]{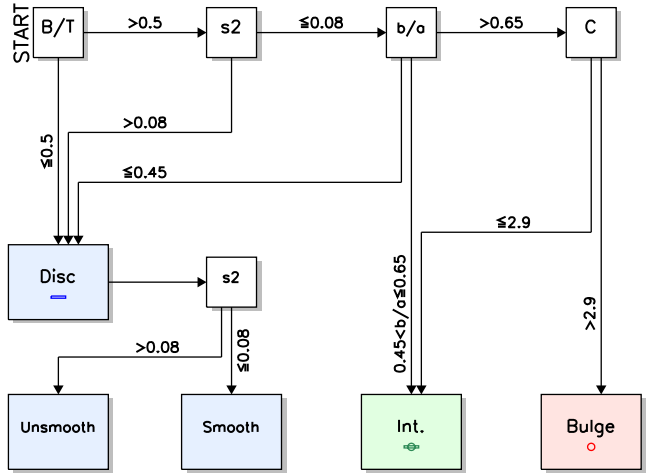}
\caption{A schematic diagram of the automated method. All 984 galaxies in our sample can be classified as automated bulges, intermediates, smooth discs, or unsmooth discs. The numerical results are tabulated in Tables~\ref{bulgedisc_clean} and ~\ref{features}.}
\label{flowchart}
\end{figure}

\noindent This logic is depicted as a flowchart in Figure~\ref{flowchart}. These are the automated classifications used in the remainder of the paper. 

It should be noted that the classifications of automated discs and intermediates depend somewhat on the precise order in which the cuts are applied. The above order was chosen so as to maximize the \textit{purity} of the automated samples. For example, there are 34 galaxies for which the order of the $s2$ and $C$ cuts would change their automated classifications. This group ($B/T > 0.5$, $s2 > 0.08$, $C\le2.9$), which consists of 1 visual bulge, 7 visual intermediates, and 27 visual discs, is placed in the automated disc sample if the $s2$ cut is applied first and in the automated intermediate sample if the $C$ cut is applied first. Since the majority of these galaxies are visual discs, the preferred order is $s2$ first.

Similarly, there are 4 visual bulges, 26 visual intermediates, and 24 visual discs that satisfy the criteria $s2 > 0.08$, $0.45<b/a\le0.65$; this region is designated to be part of the automated disc sample by making the $s2$ cut first. And lastly, there are one visual intermediate and 7 visual discs that satisfy the criteria $b/a\le0.45$, $C\le2.9$; this region is designated to be part of the automated disc sample by making the $b/a$ cut before the $C$ cut. This gives the best possible order as $s2$, $b/a$, then $C$.

The final comparison between visual and automated types is shown in Table~\ref{bulgedisc_clean}, which repeats Table~\ref{bulgedisc} using the improved automated bulge, disc, and intermediate classifications rather than using only $B/T$. Comparison with Table~\ref{bulgedisc} shows that the purity of the automated bulge sample is significantly improved over that of the high-$B/T$ sample, from $46.3 $ per cent to $73.0 $ per cent. The purity of the automated disc sample is $\sim10$ per cent lower than that of the low-$B/T$ sample, but its completeness is $\sim29$ per cent higher. With the additional parameters $s2$, $C$, $b/a$, the purities of both the automated bulge and disc samples are $\sim70 $ per cent; the completenesses are 75 per cent and 83 per cent, respectively. Of the 69 brightest visual discs ($BC=3, M_r\le-22$) discussed in \S\ref{prelimconc}, 60 (87 per cent) are correctly classified as automated discs. Indeed, 42 of them satisfy the initial condition of $B/T\le0.5$.

\begin{table}
\caption{Visual vs. Final Automated Bulge and Disc Classifications.}
\begin{tabular}{lcccc}
\hline
Visual & \multicolumn{3}{c}{Automated} & Completeness\\ 
  & Bulge   & Int. & Disc    & \\
\hline
Bulge ($BC = 1$)  &    \textbf{260}  &  39  &  47  &  75.1\\
Int. ($BC = 2$)  &  68  &    \textbf{70}  &  99  &  29.5\\
Disc ($BC = 3$)  &  28  &  41  &    \textbf{332}  &  82.8\\
\hline
Purity  &  73.0  &  46.7 &  69.5\\
\hline
\label{bulgedisc_clean}
\end{tabular}
Same as Table~\ref{bulgedisc}, but using the final, refined automated classification recipe described in the text.
\end{table}

In summary, the agreement between the visual and automated methods appears to be quite good. Furthermore, the disagreement between methods is rarely more than one type. That is, the contamination across types (i.e., $BC=3$ galaxies in the automated bulge sample or $BC=1$ galaxies in the automated disc sample) is $<10$ per cent in each automated sample. Thus, we are confident that the automated classifications can be extrapolated to determine the morphologies of a large sample that is not inspected visually, provided that sample has the same basic selection parameters as this well-tested sample.

\subsubsection{Disagreements between visual and automated classifications}\label{disagreements}
We now discuss cases where the visual and automated methods disagree in order to understand the sources of the discrepancies. First, we consider the visual discs that are classified as automated bulges. Many have low surface brightness discs, which points to a subtle difference between the two classification schemes. For the visual scheme, the criterion is that a disc be present and dominate the light in the \textit{outer} regions. For the automated scheme, the criterion is that the disc contain more of the galaxy's \textit{total} light than the bulge. In other words, the light of the galaxy may be truly bulge-dominated in terms of the total light, but the presence of a faint disc causes it to be classified as a visual disc. 

Another possible source of discrepancy is that GIM2D fits are performed on $r$-band images of the galaxies, while the visual classification is done using multi-band color images. In the $r$-band the bulge component may contribute more light, boosting $B/T$. Whatever the case may be, we conclude that the automated bulges are definitely failing to find some visual discs, but the actual amount of disc light in these interesting objects may be low and needs to be quantified via more careful modeling in the future. 

In the reverse case -- visual bulges that are classified as automated discs -- some have non-smooth features (such as dust, a double nucleus or a superposed star), which cause the galaxy to have $s2 > 0.08$. Others are visual bulges that have $B/T\le0.5$ (i.e., automated discs). Their images support the visual classification, and one possibility is that these galaxies may simply be the error tail of the $B/T$ distribution, as most of them fall near the $B/T=0.5$ boundary; only three visual bulges have $B/T\le0.4$. 

\begin{figure*}
\includegraphics[width=0.49\textwidth]{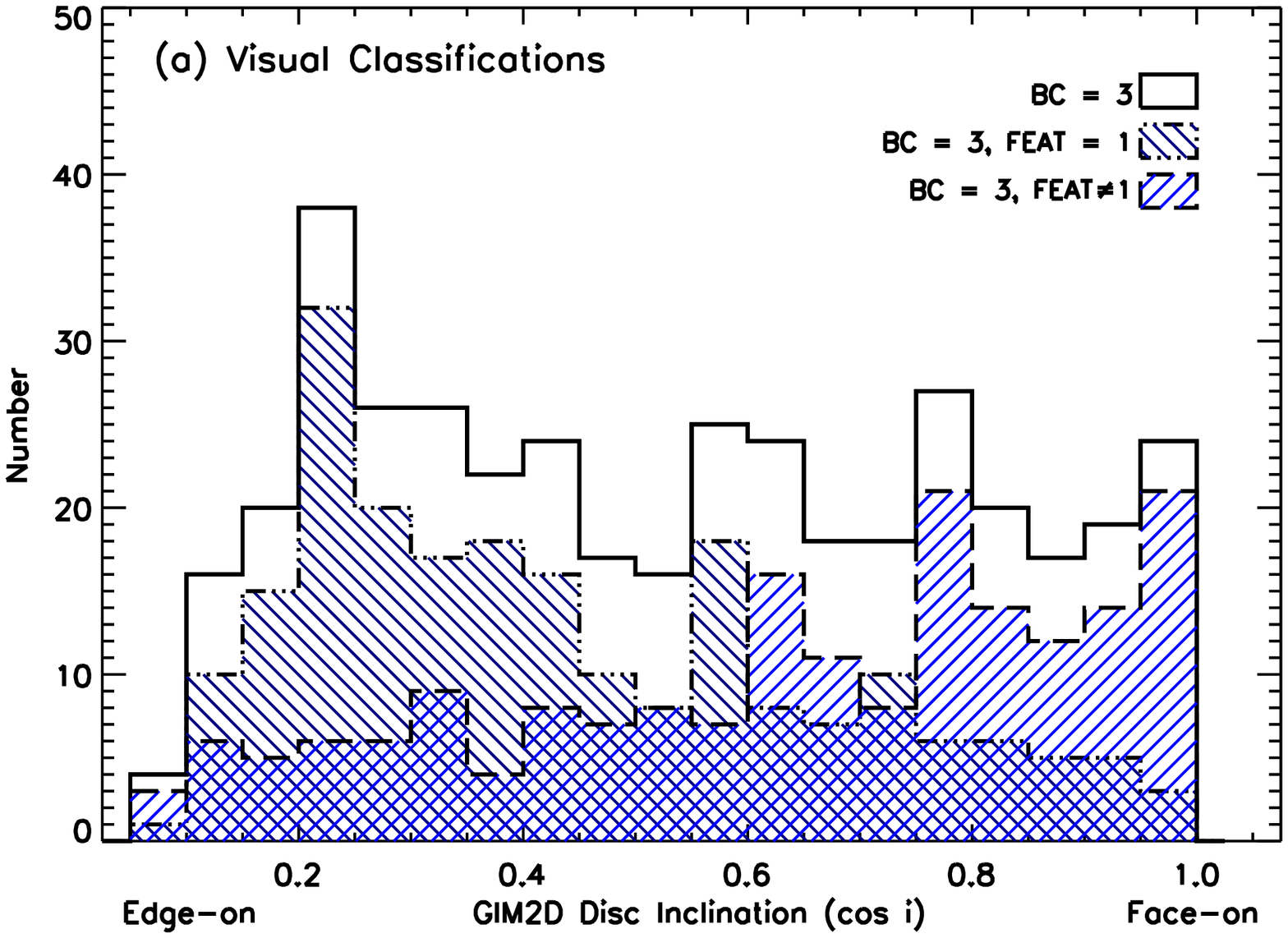}
\includegraphics[width=0.49\textwidth]{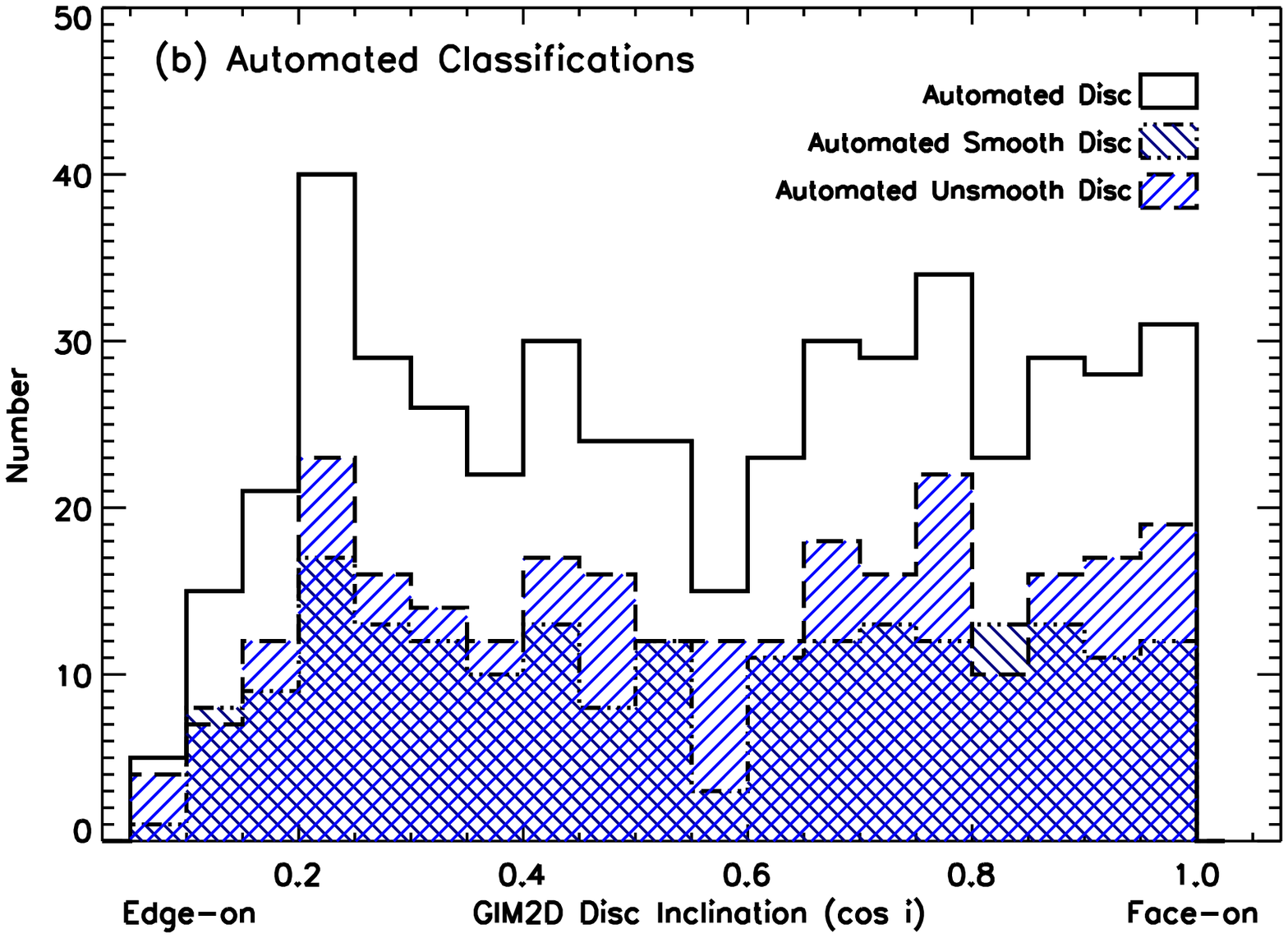}
\caption{Histograms of GIM2D disc inclination angles of (a) visual discs ($BC=3$) and (b) automated discs. (a) The distribution of visual disc inclinations indicates a slight bias for finding visual discs at higher inclinations. In an unbiased sample, we expect the distribution of cos $i$ to be flat. Above cos $i>0.3$, the observed distribution is nearly flat, but there are slightly more edge-on than face-on discs. A KS test suggests that the bias exists but is small. A stronger bias is that visual unsmooth discs are much more likely to be face-on ($FEAT\neq1$) than visual smooth discs ($FEAT=1$). (b) Inclination biases in automated discs are smaller. The overall distribution as well as the automated smooth ($s2\le0.08$) and unsmooth ($s2>0.08$) samples have much flatter distributions than the corresponding visual samples. GIM2D is thus able to identify discs more uniformly than the visual method and can spot non-smooth features at all inclinations.}
\label{inc_bias}
\end{figure*}

We also consider the possibility that a bias in the visual classification scheme causes face-on discs (hidden within the bulge component) to be more difficult to identify by eye. Such galaxies would be more likely to be classified visually as bulges or intermediates. Figure~\ref{inc_bias}(a) shows the distribution of disc inclinations (cos $i$) determined by GIM2D. The upper black line shows the distribution for all visual discs. The lack of galaxies at cos $i = 0.0$ and the spike at cos $i\sim0.2$ appear to be an artefact from the GIM2D fits that is seen in the distribution of the total sample, so it should not be taken as evidence for a bias in finding discs. This feature is likely due to the fact that a galaxy cannot be perfectly edge-on because real discs are not infinitely thin as GIM2D assumes so that truly edge-on discs are piled up at cos $i\sim0.2$. Above cos $i>0.3$, there does appear to be a slight decline in the number of visual discs as cos $i$ increases. A KS test comparing the distribution at cos $i>0.3$ to a flat distribution yields a probability of 3.9\%. We conclude from this that there is indeed a slight bias against visually identifying face-on discs; the effect is statistically significant, but small. 

\begin{figure*}
\includegraphics[width=0.49\textwidth]{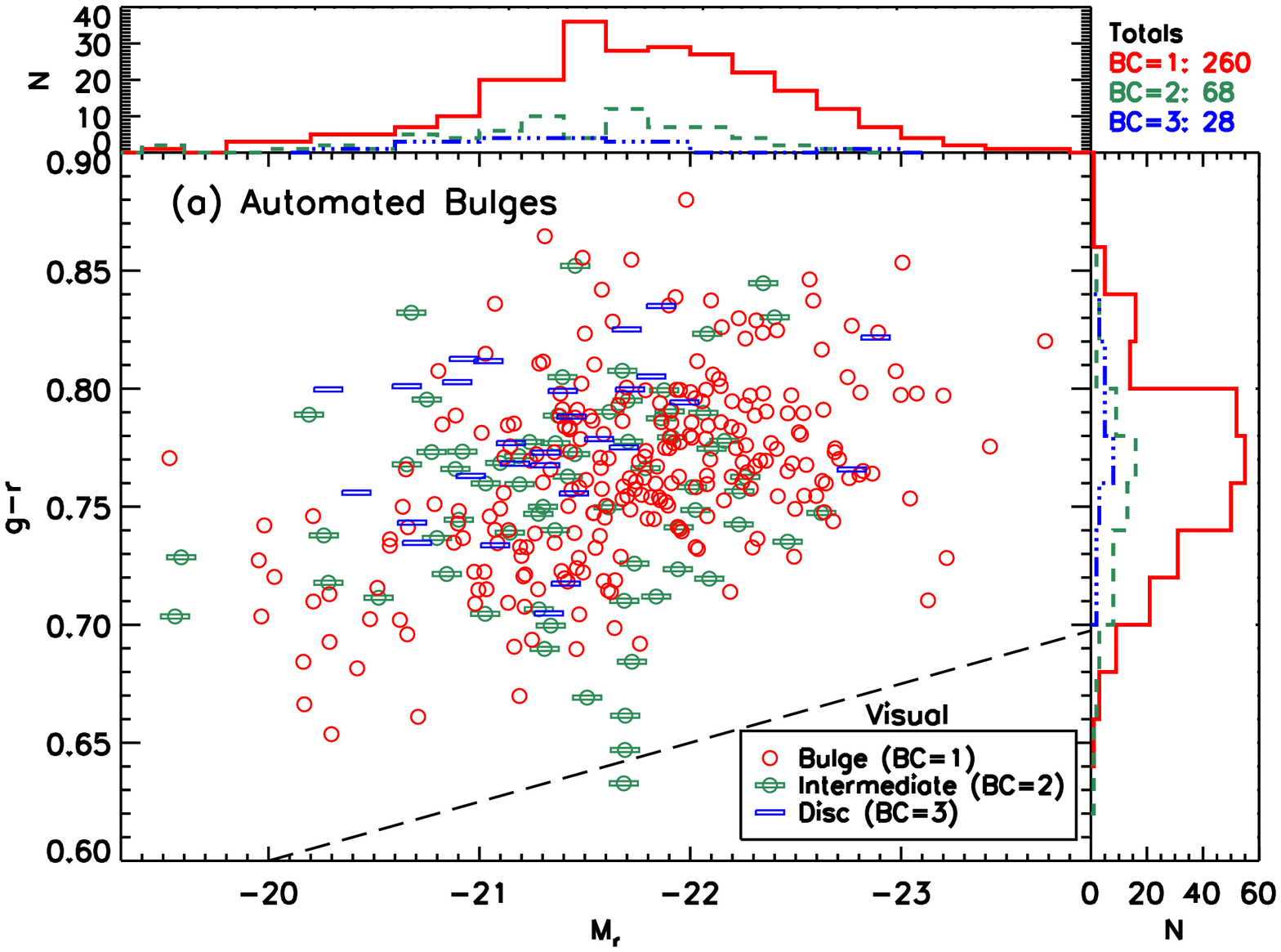}
\includegraphics[width=0.49\textwidth]{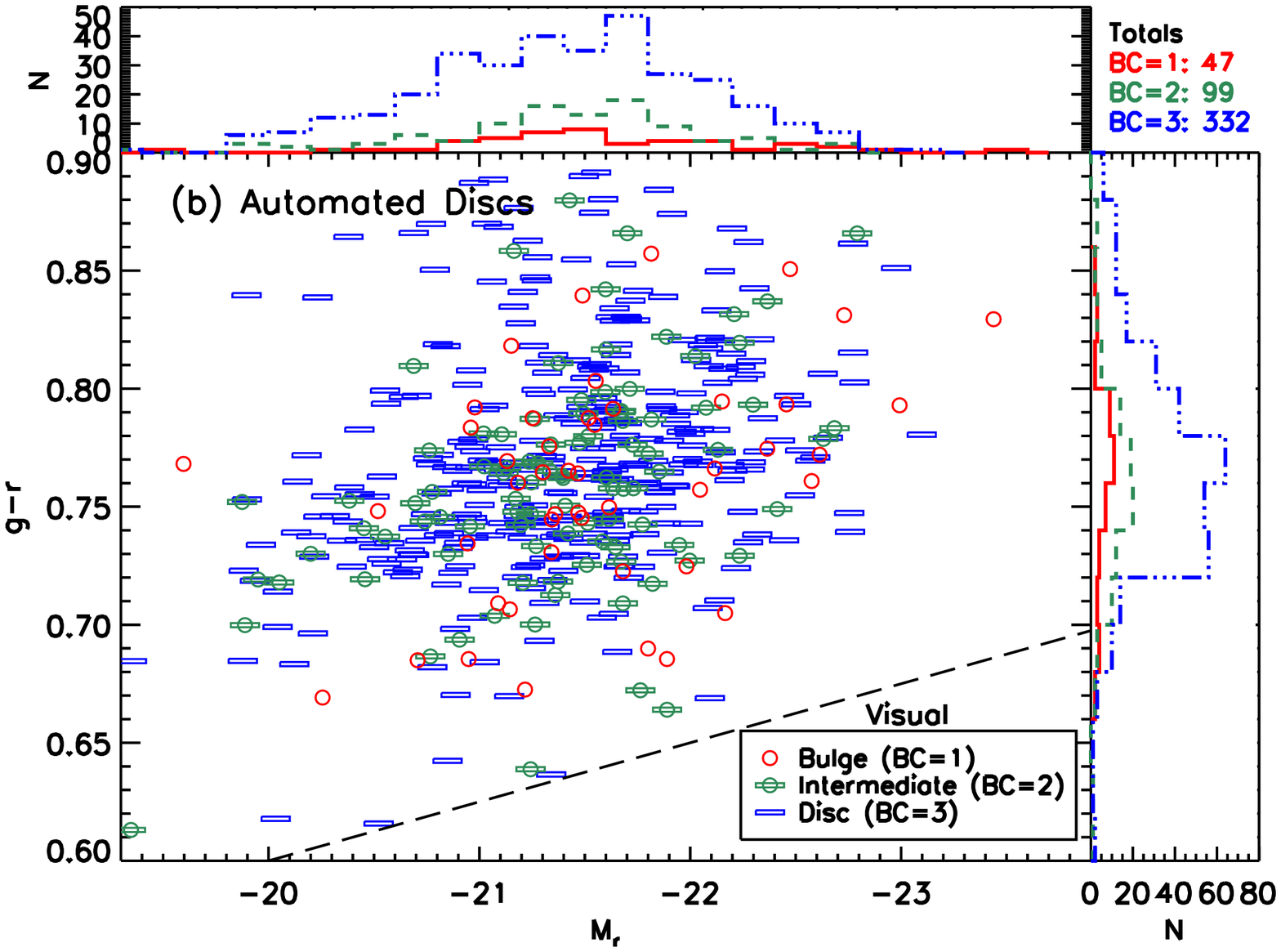}
\caption{Same as Figure~\ref{cmd} but divided according to automated type: (a) the automated bulge sample of 356 galaxies and (b) the automated disc sample of 478 galaxies. Comparison with the histograms of Figure~\ref{cmd} reveals that discrepant galaxies in both automated samples may tend toward fainter magnitudes, but the effect is more pronounced for visual discs classified as automated bulges (blue bars, panel a).}
\label{auto_cmd}
\end{figure*}

Figure~\ref{auto_cmd} explores misclassifications as a function of absolute magnitude by plotting separate color-magnitude diagrams of (a) the automated bulge sample and (b) the automated disc sample, with symbols indicating the visual classifications in both panels. The discrepant galaxies in both automated samples may tend toward fainter magnitudes, but the effect is more pronounced for visual discs classified as automated bulges (blue bars, panel a).

\subsection{Smooth vs. unsmooth discs}\label{smoothunsmooth}
To study the properties of quenched galaxies in a large sample, we would like to be able to identify star-forming contaminants in our sample. Visually these are unsmooth discs with $FEAT\neq1$ and make up 23 per cent of our total sample. In this section, we will show that we are able to use GIM2D smoothness $s2$ to reliably remove star-forming contaminants from future samples using only their automated morphological classifications. The following analysis applies only to the automated disc sample described above. We use the \textit{automated} discs because in the future, in the absence of visual inspection, we will be able to use only the automated parameters.

\begin{figure}
\includegraphics[width=0.5\textwidth]{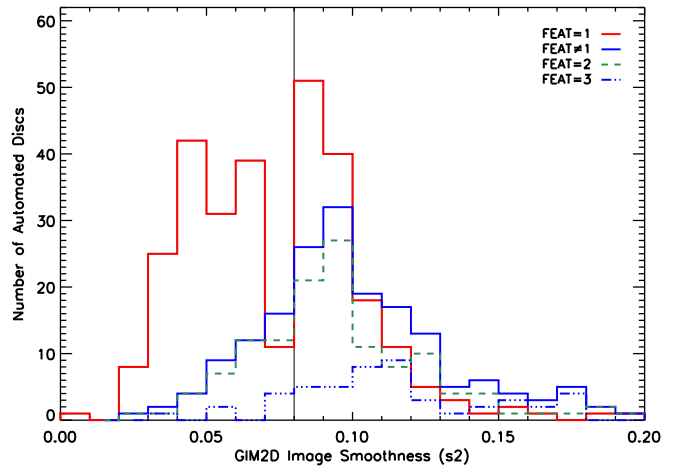}
\caption{Histograms showing the correlation between GIM2D smoothness $s2$ and the visual classifications $FEAT$ for automated discs. Most automated discs with $s2\leq0.08$ are visually smooth (red, $FEAT=1$), while most automated discs with $s2>0.08$ are visually unsmooth (solid blue, $FEAT\ne1$). The numerical results are tabulated in Table~\ref{features}.}
\label{feathist}
\end{figure}

Figure~\ref{feathist} shows the correlation between $FEAT$ and $s2$. The distribution of visual smooth discs ($FEAT=1$, red line) is centred at a lower value of $s2$ than visual unsmooth discs ($FEAT\neq1$, blue solid line). We therefore divide the sample into two populations, one with low smoothness parameter ($s2\leq0.08$), which includes half of the visual smooth discs, and one with high smoothness parameter ($s2>0.08$), which includes most of the visual unsmooth discs. The numbers of each visual type falling on both sides of the $s2=0.08$ cut are listed in Table~\ref{features}. The purity of the automated smooth disc sample is high (78 per cent), but the purity of the automated unsmooth disc sample is rather low (49 per cent). The low purity of the automated unsmooth disc sample may be due to a bias in the visual classifications.

\begin{table}
\caption{Visual vs. Automated Smooth and Unsmooth Classifications (Automated Discs Only).}
\begin{tabular}{@{}l@{}ccc@{}}
\hline
Visual & \multicolumn{2}{c}{Automated} & Completeness\\
  & $s2\leq0.08$    & $s2>0.08$    & \\
\hline
Smooth ($FEAT = 1$)  &   \textbf{158}  &  141  &  52.8\\
Unsmooth ($FEAT = 2$) &  38  &   \textbf{90}  &  70.3\\
Unsmooth ($FEAT = 3$) &  7  &   \textbf{44}  &  86.3\\
\hline
Purity  &  77.8 &  48.7\\
\hline
\label{features}
\end{tabular}
Same as Table~\ref{bulgedisc}, but comparing visual and automated smooth vs. unsmooth discs.
\end{table}

As in the previous section, we examine cases where the two methods disagree. Many visual unsmooth discs ($BC=3, FEAT\ne1$) that are classified as \textit{automated smooth discs} ($s2\leq0.08$) have rather weak features, and nearly all have features that are reflection symmetric. Because the smoothness parameter is in part a measure of asymmetry, it is plausible that a symmetric, face-on spiral or a disc with a bar may be featureless by the automated criteria. One possible way to check this quantitatively is to treat the total residual ($R_T$) and asymmetric residual ($R_A$) separately, though for simplicity, we do not attempt it here because these galaxies make up less than five per cent of our sample.

For the reverse case, visual smooth discs ($BC=3, FEAT=1$) that are classified as \textit{automated unsmooth discs} ($s2>0.08$), there is a strong tendency for the galaxies to be edge-on. This bias is understandable: with the exception of dust, which is seen almost exclusively in edge-on discs, features are more likely to be picked out visually if a galaxy is face-on. In a highly inclined disc the light is projected on to a smaller area so it is more difficult to see, by eye, a contrast in brightness between a spiral arm or a bar and the rest of the light in the galaxy. It seems clear that the visual classifications are not spotting all features in highly inclined galaxies.

This is shown more quantitatively in Figure~\ref{inc_bias}(a), where the blue hatched histograms show the disc inclination distributions for visual smooth and unsmooth discs. As in the previous discussion of inclination bias (\S\ref{disagreements}), the distribution of cos $i$ should be flat because all inclinations are equally likely. In Figure~\ref{inc_bias}(a), however, the number of visual smooth discs increases steadily as the inclination becomes closer to edge-on (cos $i=0$), while the number of visual unsmooth discs increases steadily as the inclination becomes closer to face-on (cos $i=1$). The increase is smooth and regular, which is consistent with a real bias in identifying features in inclined discs.

The behavior of \textit{automated} smooth and unsmooth discs is very different, as shown in Figure~\ref{inc_bias}(b). The distributions are much flatter, which may indicate that GIM2D is much better than visual classification at finding substructure in edge-on discs. With this in mind, we re-examine the results shown in Table~\ref{featnum}, restricting attention in the rightmost columns to face-on galaxies with cos $i>0.8$. We see that visual features ($FEAT\ne1$) are much more common in face-on discs than when considering all of the visual discs (\S\ref{prelimconc}). The bias against finding edge-on visual unsmooth discs may explain the apparent low purity of the automated unsmooth disc sample (Table~\ref{features}). That is, these galaxies may really possess features but the visual method fails to detect them.

Based on the results of this section, 275 star-forming contaminants (automated unsmooth discs) can be removed from our original sample of 984 using the automated classification scheme. Of the remaining 709 quenched galaxies, only 67 (9.4 per cent) are visual unsmooth discs. Figure~\ref{inc_bias} indicates that many of the visual smooth discs have underlying features that are difficult to pick out by eye because of inclination effects. This is not a problem, however, for the automated method, so future samples that are selected using the automated parameters will not be severely contaminated. 

\begin{figure*}
\includegraphics[width=0.49\textwidth]{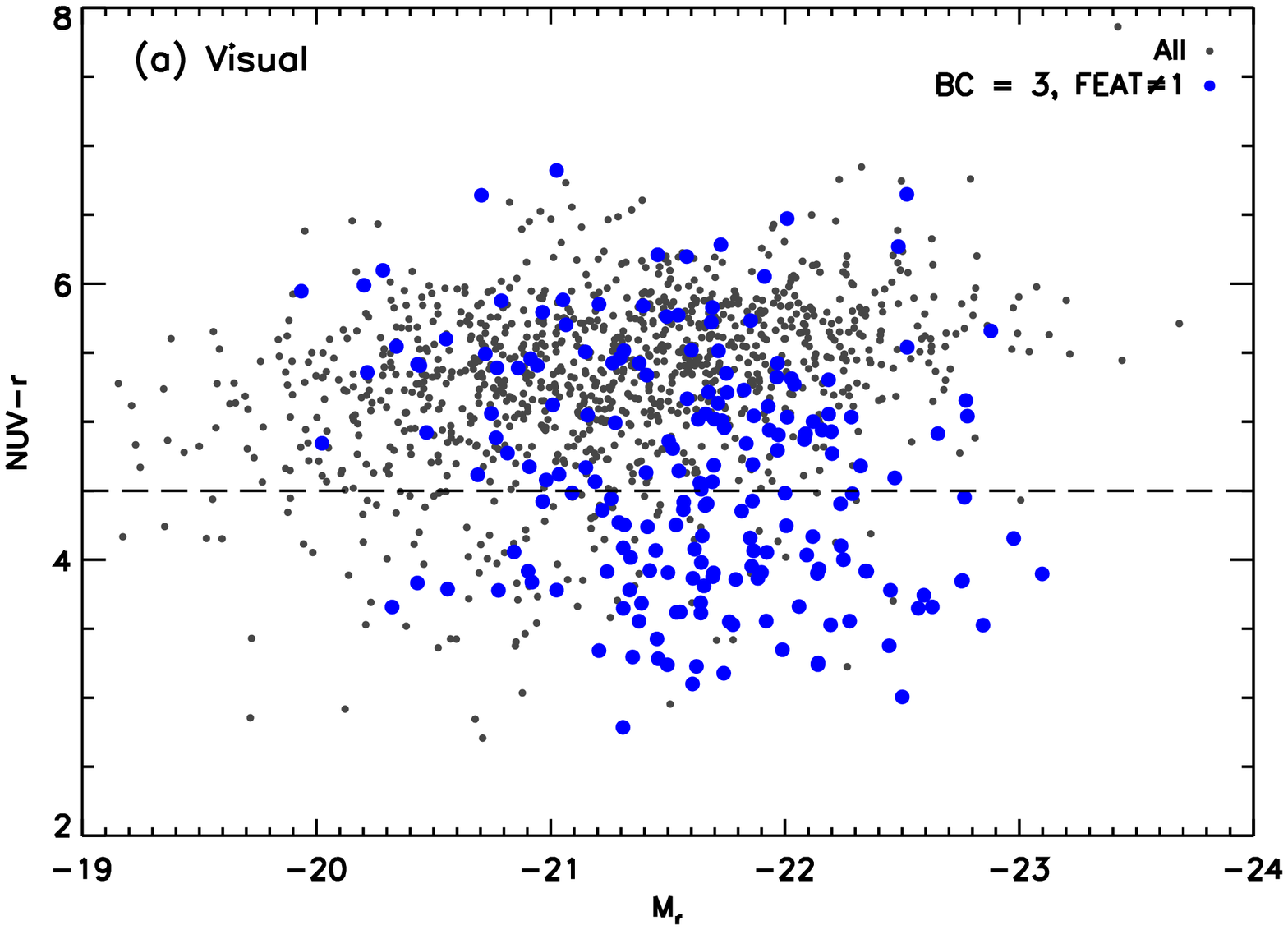}
\includegraphics[width=0.49\textwidth]{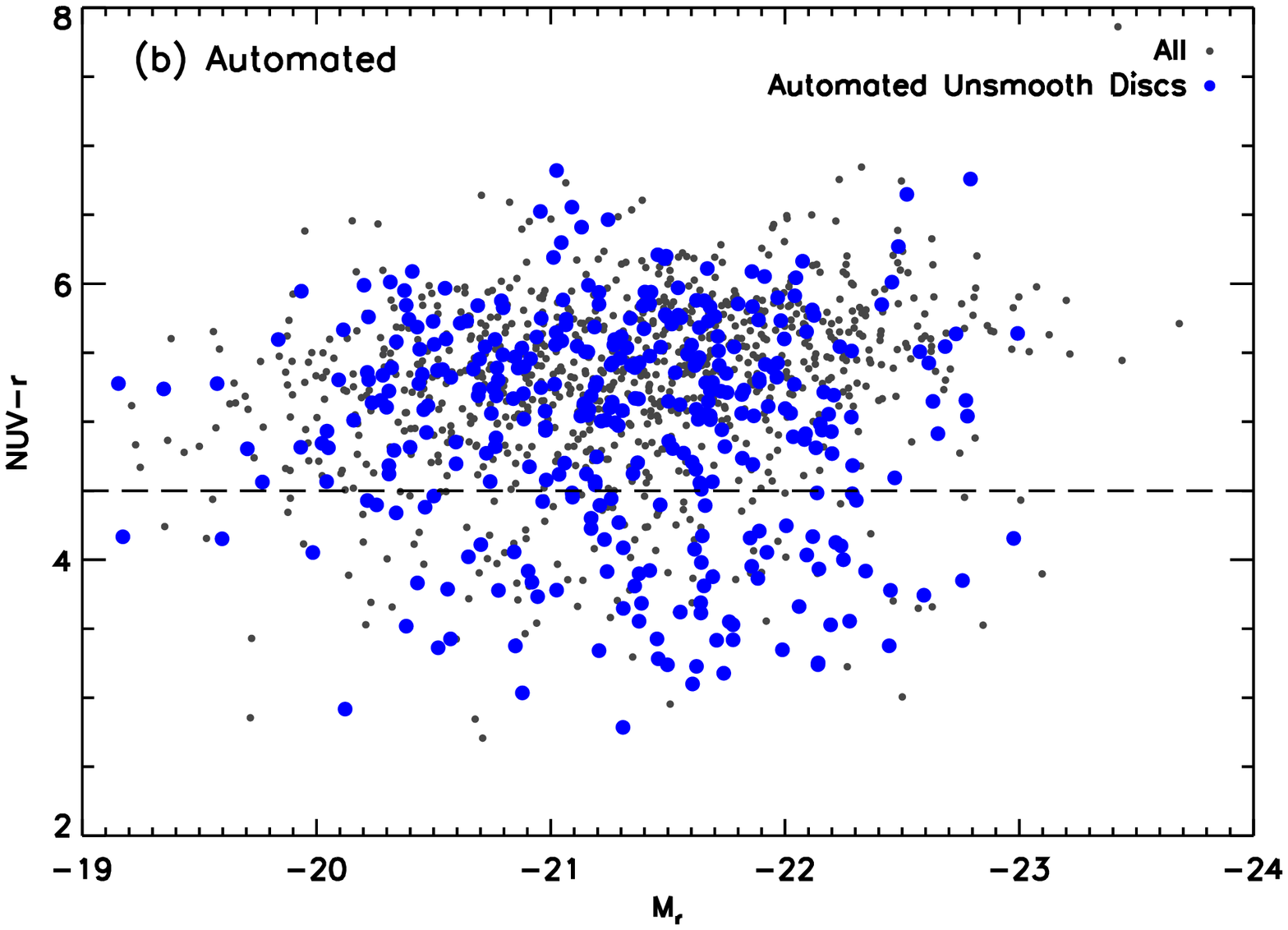}
\caption{Color-magnitude diagram using GALEX $NUV-r$. The bluest galaxies are likely to be star-forming. The horizontal line indicates a rough cut below which we consider galaxies to be contaminants. The blue symbols indicate the galaxies that are unsmooth discs (i.e., star-forming) using the (a) visual and (b) automated methods. Either method athes roughly half of all star-forming galaxies below the dashed line, leaving a quenced sample that is $\sim 90$ per cent pure (see text). Our analysis shows that using the morphology to identify star-forming galaxies is an effective way to remove contaminants from the quenched sample.}
\label{cmd_nuv}
\end{figure*}

As a final check of how severe residual contamination by star-forming galaxies is in our sample, we compare our sample of unsmooth discs to galaxies that have the bluest GALEX $NUV-r$ colors. As GALEX photometry is not available for all SDSS galaxies, we do not want to use it in selecting our sample. Nevertheless, the UV photometry can serve as a valuable check on the presence of young stars for the smaller sample of galaxies presented here. Figure~\ref{cmd_nuv} shows a CMD of our sample using the GALEX $NUV-r$ color. There is an obvious tail of blue galaxies with colors that suggest that they are star-forming galaxies. This tail is largely absent in the $g-r$ CMD of Figure~\ref{cmd}, indicating that $g-r$ is not sensitive enough to young stars for our purposes. This is true even with improved photometry from GIM2D. In the following, we will compare galaxies with $NUV-r\le4.5$ with the unsmooth discs; these are indicated in blue in Figure~\ref{cmd_nuv} for both the (a) visual and (b) automated classifications.

In the original sample of 984 galaxies, 158 have $NUV-r\le4.5$; 90 of these are visual unsmooth discs and 78 are automated unsmooth discs, which suggests that roughly half of blue galaxies can by identified morphologically using either method. In the remaining sample of 709 nominally quenched galaxies, 80 have $NUV-r\le4.5$. If we take all of these galaxies as being star-forming, about 10 per cent of the \textquoteleft quenched' sample is made up of star-forming contaminants. This should be an upper limit, as some of the galaxies with $NUV-r\le4.5$ may simply be the tail of the red sequence.

Based on the above analysis, we believe that approximately half of the star-forming contaminants can be effectively removed by identifying them morphologically. In the remaining sections, we analyze only the \textit{quenched} sample, that is, bulges, intermediates, and \textit{smooth} discs. The 275 star-forming contaminants identified by their unsmooth morphology in this section are not included in the following analysis. We also examine whether the remaining blue contaminants -- those that were not identified by their unsmooth morphology -- affect on the result.

\section{Morphologies vs. Structural Parameters using Both Methods}\label{results}
Now that both visual and automated classifications are in place, the remainder of this paper employs them in a first reconnaissance of how the morphologies of quenched early type galaxies vary versus galaxy structural parameters. Because bulges and discs may have experienced different quenching mechanisms, the frequencies of each type as a function of various structural parameters will provide constraints for quenching theories. \textit{In this section, \textquoteleft discs' refers only to non-starforming smooth discs -- unsmooth discs (i.e., blue contaminants) have been removed using the criteria outlined in the previous section.} We present some preliminary results for our sample of 709 quenched galaxies and also assess the degree to which our conclusions depend on which method is used. By and large, we find that the two methods agree well in a statistical sense, which shows that our automated method can be used with confidence for a larger sample to examine various scaling relations for bulges and discs separately. 

The size parameters used here are $r$-band absolute magnitude $M_r$, stellar mass $M_*$, velocity dispersion $\sigma$, and 50\% Petrosian radius $R_{50}$. Absolute magnitudes were derived from the model magnitudes provided by the SDSS as described in \S\ref{sdssparams}. Stellar mass measurements are taken from \citet{gal05}; it should be noted that 40 galaxies have log $M_*=-99$ (i.e., flagged as bad data values) in their catalog and are excluded in the $M_*$ histograms. The velocity dispersion is a product of the SDSS pipeline (DR6, \citealp{ade08}). The velocity dispersions used here are not aperture corrected, but the corrections are small, of order $\sim6 $ per cent \citep{gra09a}. The \textit{physical} 50 per cent Petrosian radius is calculated using the apparent size and the object's redshift.

All four of these parameters ($M_r$, $M_*$, $\sigma$, and $R_{50}$) are commonly used size indicators for early type galaxies, and previous studies have already shown that the relative fraction of bulge-dominated galaxies is greater among larger objects (\citealp{san85,mar94,mar98,mar99}). A recent paper by \citet{van09} showed that massive red objects tend to be rounder (i.e., not disc-dominated). We therefore expect to find more bulges among brighter and bigger objects. However, recent stellar population work has revealed different trends for stellar population parameters depending on which size indicator is used, and in general trends with $\sigma$ are the sharpest (\citealp{gra08,gra09a,gra09b,gra10}). We are interested in looking for differences in the bulge/disc frequencies as a function of the size parameters used here: will $\sigma$ again give the sharpest trend? Finally, since $\sigma$ and $R_{50}$ present the Fundamental Plane nearly face on, we can use the present sample to attempt a first mapping of bulge/disc frequencies directly \textit{on} the Fundamental Plane.

\begin{figure*}
\includegraphics[width=1.0\textwidth]{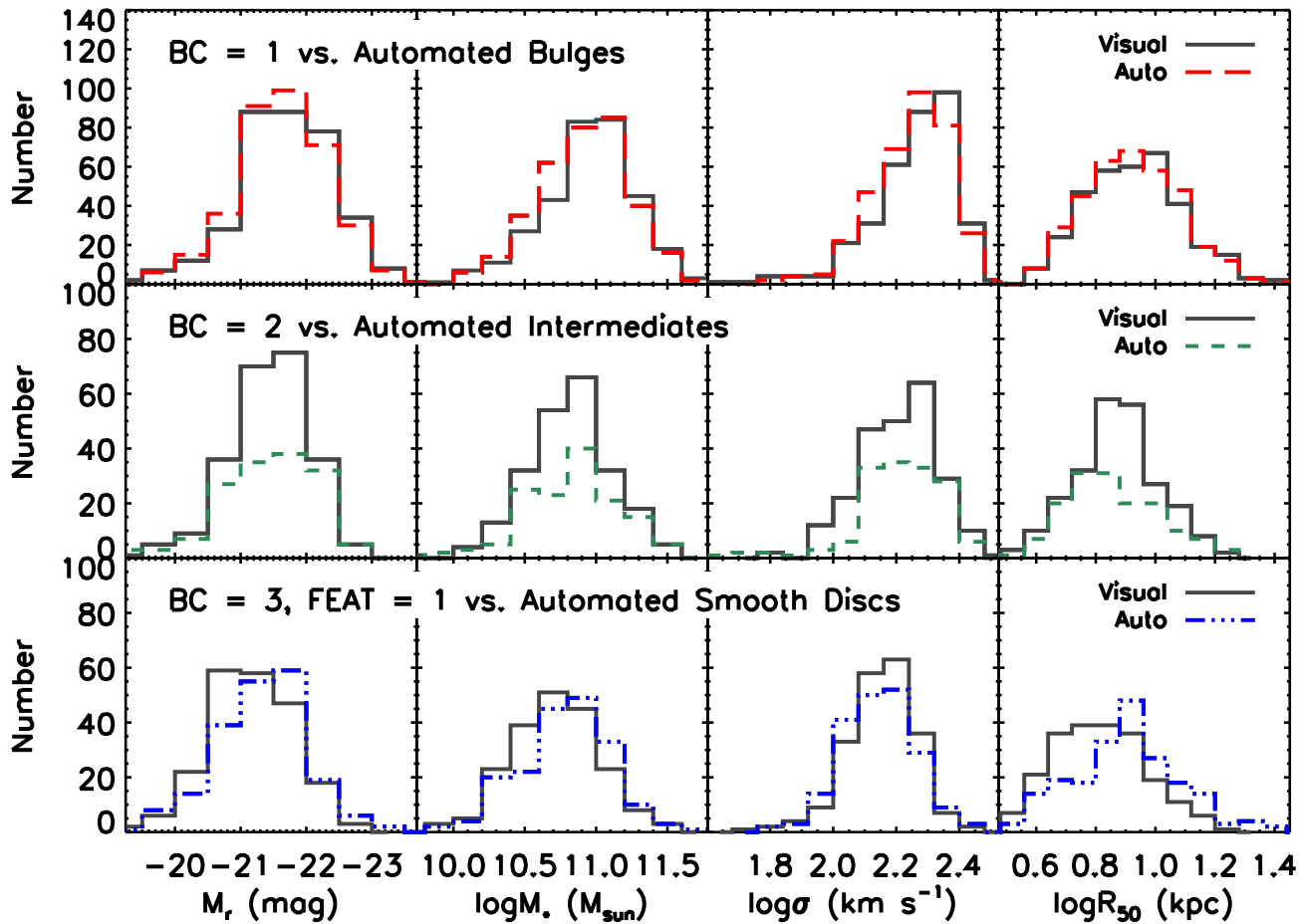}
\caption{Comparison of the $r$-band magnitude $M_r$, stellar mass $M_*$, velocity dispersion $\sigma$, and 50 per cent Petrosian radius $R_{50}$ distributions of bulges, intermediates, and smooth discs using visual (gray) and automated (colored) classifications. The agreement between the methods is reasonably good for bulges (top) and discs (bottom), but worse for intermediates (middle), consistent with the results of \S\ref{autoscheme} and Table ~\ref{bulgedisc_clean}.}
\label{hist_compare1}
\end{figure*}

Figure~\ref{hist_compare1} compares the visual and automated methods directly by overplotting the two distributions for each type and size parameter. Visual classifications are plotted in gray, while the automated classifications are plotted in color. Bulge distributions (top row) agree extremely well in both shape and number of galaxies. Disc distributions (bottom row) also agree well with the possible exception that there are fewer small discs in the automated disc sample.

This can be explained by noting that the objects that fall in opposite visual and automated classes (that is, visual bulges in the automated disc sample and vice versa) roughly follow the same distributions as their parent populations. Visual bulges occupy a much wider range in magnitude and go to brighter magnitudes than the visual discs. This can be seen by looking at the color-magnitude diagrams of the automated samples (Figure~\ref{auto_cmd}). Consequently, visual disc contaminants in the automated bulge sample are of small or intermediate size, but visual bulge contaminants in the automated disc sample can range from very small to very large. This has the effect of increasing the number of outliers in the automated disc sample while decreasing the number of outliers in the automated bulge sample.

The \textit{shapes} of the distributions for intermediate galaxies (middle row) also agree well, but fewer galaxies overall are classed as intermediates using the automated method compared to the visual method (a 37 per cent decrease, from 237 to 150). This is not a fundamental discrepancy, as the relative numbers in the different automated classes could be easily adjusted by moving the classification boundaries slightly in Figures~\ref{bf_s2},~\ref{bf_abr}, and~\ref{bf_conc}. We elect to leave the boundaries where they are for now and defer any changes to future work, if needed.

The summary of numbers in Table~\ref{bulgedisc_clean} suggests that the automated intermediate sample contains roughly the same numbers of visual bulge and disc contaminants. Furthermore, our previous attempt to more finely classify intermediates (\S\ref{visualscheme}) yielded approximately the same number of bulgy and discy intermediates. Finally, the most robust result of Table~\ref{bulgedisc_clean} is that pure bulges are rarely confused with pure discs but that intermediates are less reliably identified.

\begin{figure*}
\includegraphics[width=1.0\textwidth]{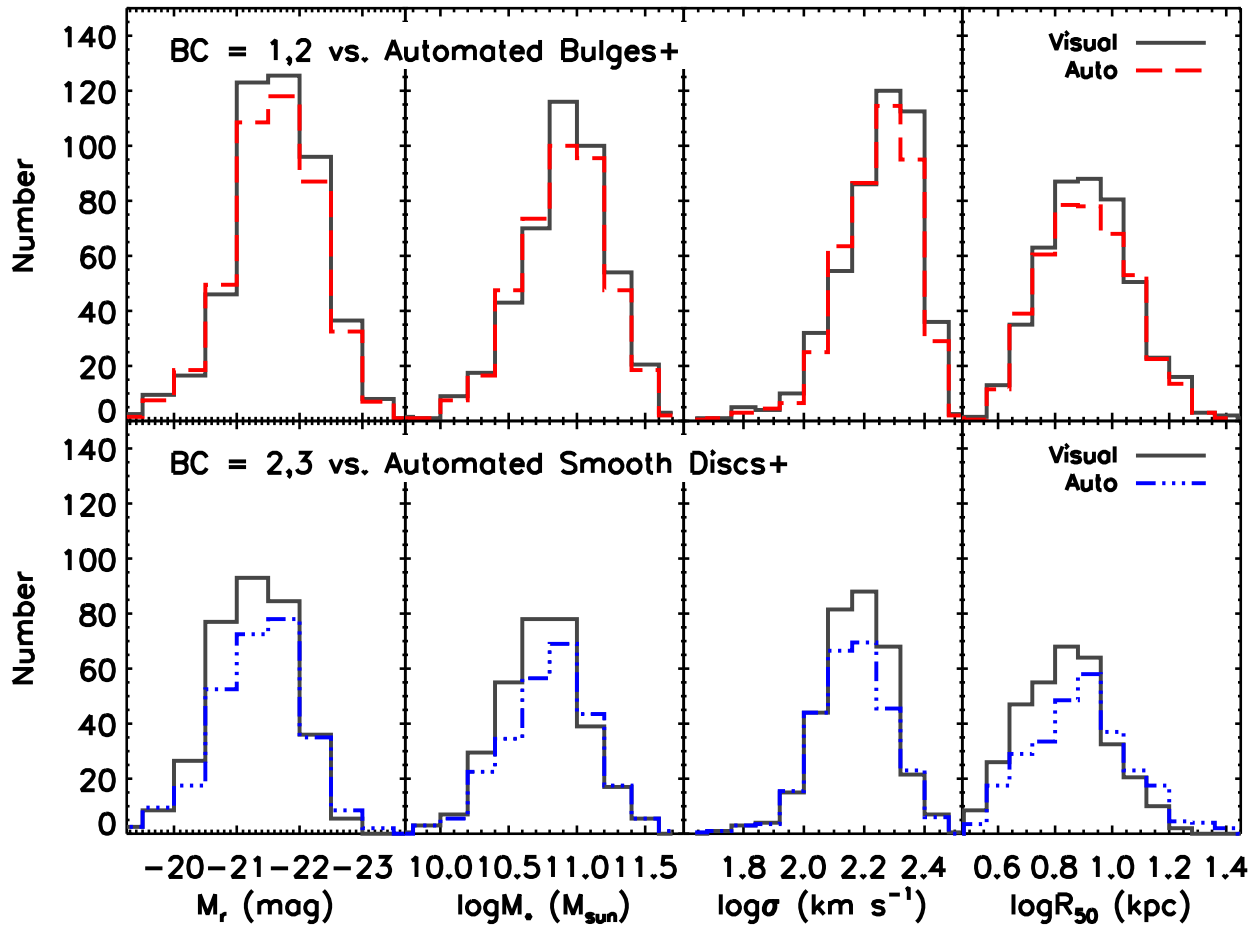}
\caption{Same as Figure~\ref{hist_compare1} but for the new samples bulges+ and discs+. When the intermediate sample is split evenly between the bulge and disc samples, the agreement in the distributions of the visual and automated samples is even better than in Figure~\ref{hist_compare1}.}
\label{hist_compare2}
\end{figure*}

With these points in mind, we abandon the intermediate classification altogether and assign half of the intermediate sample \textit{in each bin} to the bulge and disc samples. We will refer to these new samples as bulges+ (bulge sample plus half of the intermediate sample) and discs+ (disc sample plus half of the intermediate sample). The same procedure is followed for both the visual and automated classifications. Figure~\ref{hist_compare2} compares the results by overplotting the distributions from both methods for each new type. With the intermediates divided in half, the agreement between the two classification schemes is now even better.

\begin{figure*}
\includegraphics[width=1.0\textwidth]{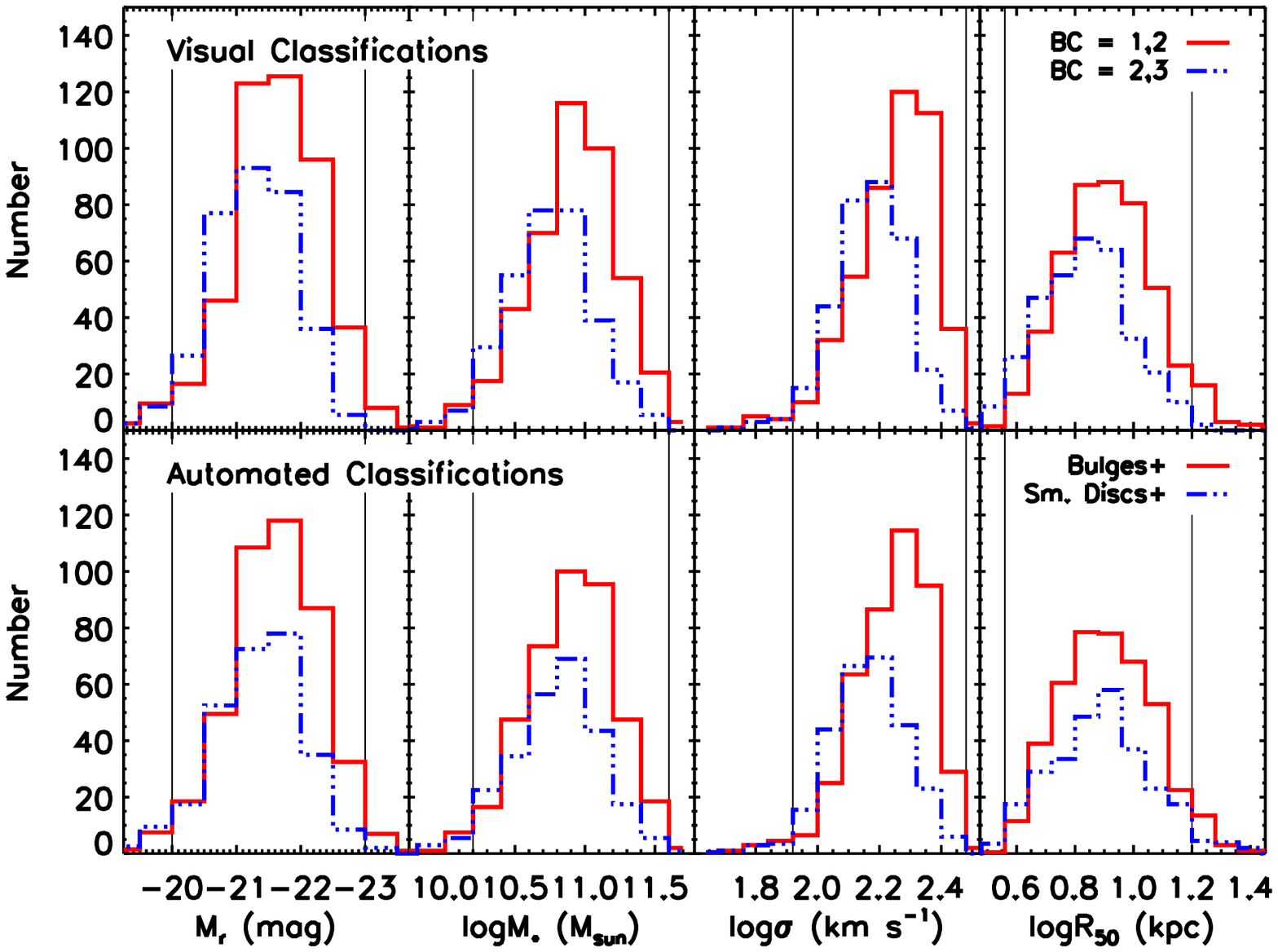}
\caption{The numbers of bulges+ (red) and discs+ (blue dash-dotted) as a function of four measures of galaxy size: $r$-band magnitude $M_r$, stellar mass $M_*$, velocity dispersion $\sigma$, and 50 per cent Petrosian radius $R_{50}$. The vertical lines indicate the boundaries within which the total number of objects in each bin is greater than 20. Bulges+ tend to be brighter, with larger $M_*$, $\sigma$ and $R_{50}$ compared to discs+.}
\label{hist_split_num}
\end{figure*}

\begin{figure*}
\includegraphics[width=1.0\textwidth]{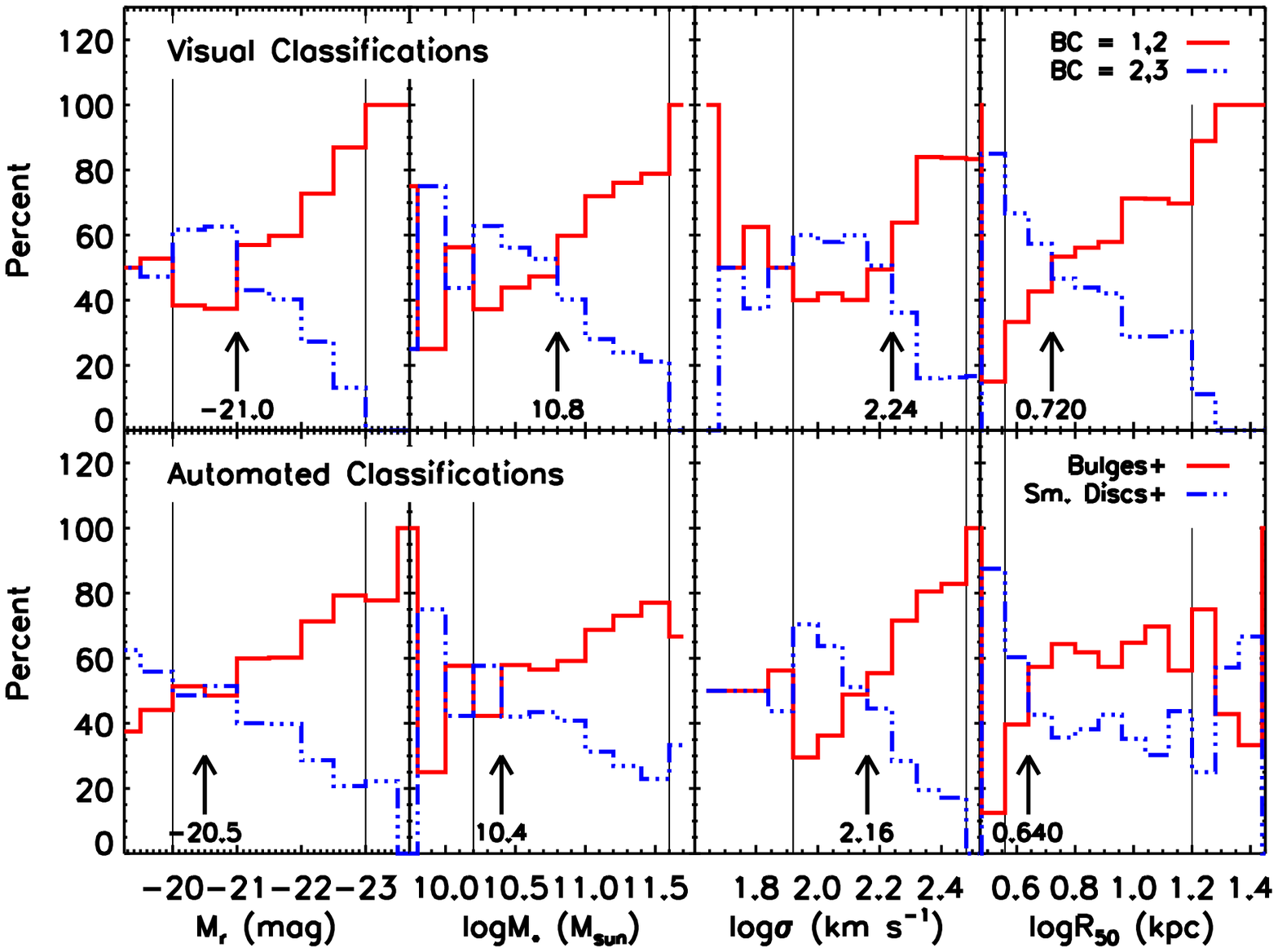}
\caption{The percentages of bulges+ (red) and discs+ (blue dash-dotted) as a function of four measures of galaxy size: $r$-band magnitude $M_r$, stellar mass $M_*$, velocity dispersion $\sigma$, and 50 per cent Petrosian radius $R_{50}$. The vertical lines indicate the boundaries within which the total number of objects in each bin is greater than 20. The \textit{crossing value} is indicated in each panel with an arrow. The general trends and crossing values agree well using both methods.}
\label{hist_split_per}
\end{figure*}

Figures~\ref{hist_split_num} and~\ref{hist_split_per} show the number and percentage, respectively, of bulges+ (red) and discs+ (blue, dash-dotted) as a function of our four size parameters. The vertical lines indicate the boundaries within which the total number of objects in each bin is greater than 20. In Figure~\ref{hist_split_per} the \textit{crossing value} in each panel is calculated and is shown in the panel with an arrow. If there are multiple crossings, the average of the first and last crossings within the reliable range is taken. A comparison of the crossing values for the visual (top panels) and automated (bottom panels) methods shows that they are in good agreement, with differences less than $\sim0.5$ dex. Varying the bin sizes does not significantly affect the crossing values or the agreement between the two methods. 

It should be noted that the trends seem somewhat stronger using the visual method because some large visual bulges are classed as discs using the automated method. This can be explained by the previous observation that most bright galaxies are bulges, so errors at the bright end are more likely to move objects from bulges to discs. It is encouraging, however, that the trends are still clearly observed using both classification methods, and even the relative strengths of the different trends are preserved.

\begin{figure*}
\includegraphics[width=0.49\textwidth]{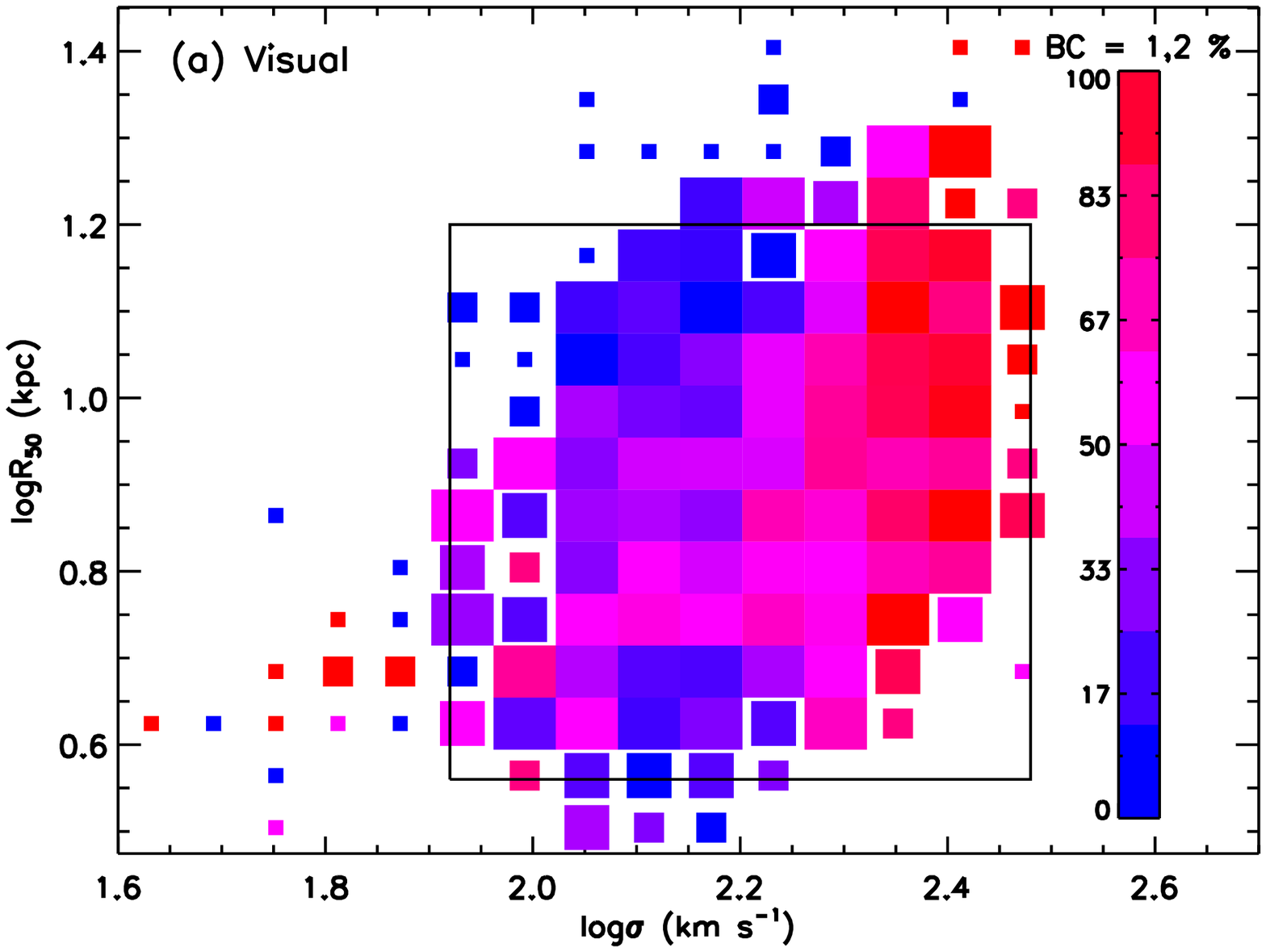}
\includegraphics[width=0.49\textwidth]{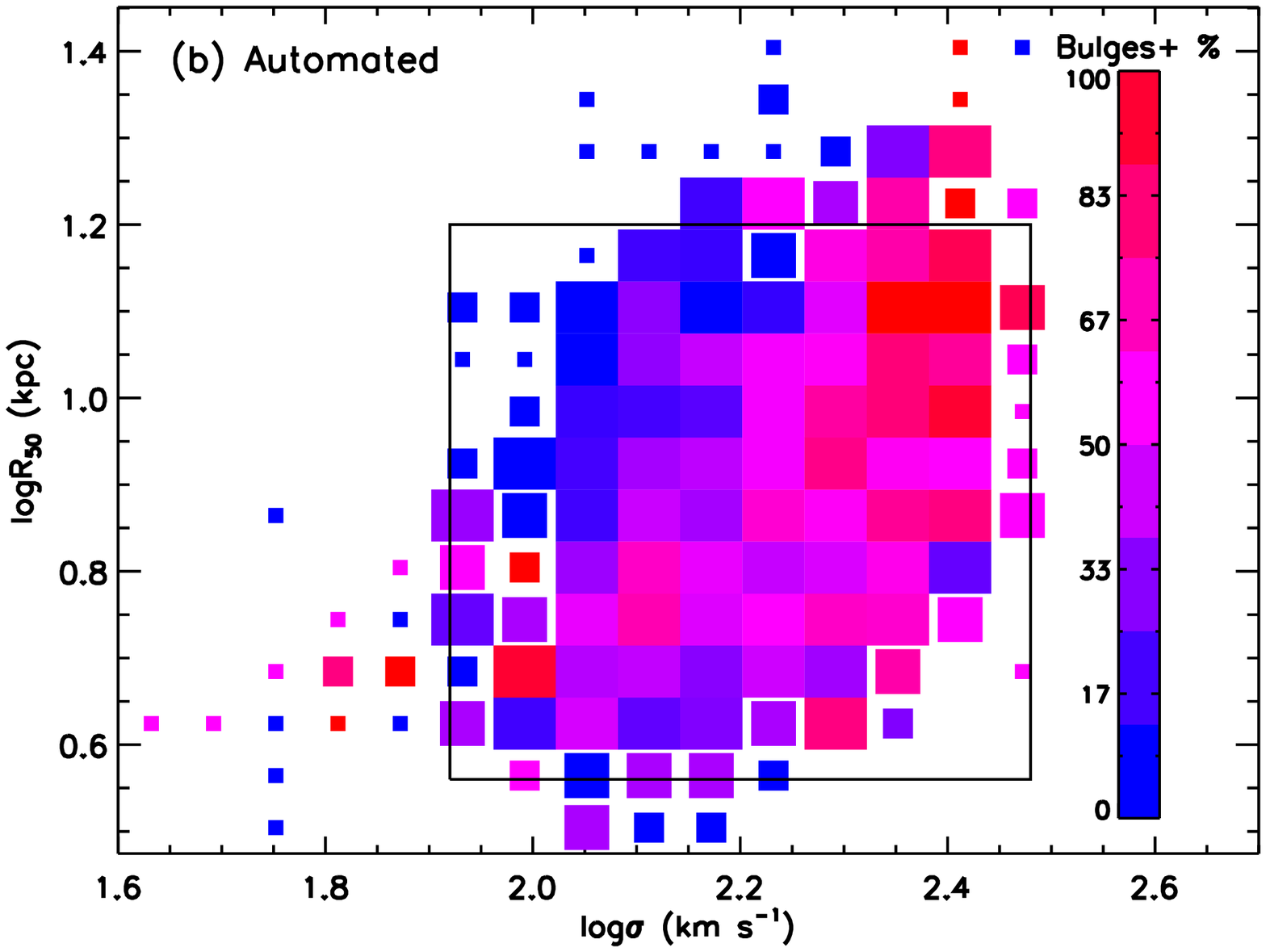}
\caption{The bulge+ frequency mapped on to the Fundamental Plane using the (a) visual and (b) automated classifications. Red indicates a bin comprised solely of bulges+, while blue indicates a bin comprised solely of discs+. Small squares indicate bins that contain fewer than four galaxies, with the size indicating the number. The black lines indicate the locations of the vertical lines in Figures~\ref{hist_split_num} and~\ref{hist_split_per}. At high $\sigma$ the sample is clearly dominated by bulges+, while at low $\sigma$ there are many more discs+. The trend in $R_{50}$ may be somewhat weaker. Again, there is good qualitative agreement between the two methods.}
\label{fundplane}
\end{figure*}

Finally, Figure~\ref{fundplane} shows the bulge+ frequency mapped onto a nearly face-on projection of the Fundamental Plane, shown for both the (a) visual and (b) automated classifications. Looking at trends in the horizontal direction (at constant $R_{50}$), bulges+ clearly dominate at high $\sigma$, while more discs+ are present at low $\sigma$. This is especially evident at the largest values of $R_{50}$. No clear trends are seen in the vertical direction (at constant $\sigma$), which may suggest that the bulge+ frequency is a stronger function of $\sigma$ than $R_{50}$. We defer a quantitative analysis of these trends to future work with a larger sample of SDSS galaxies. Though the observed trends are slightly weaker for the automated method, there is good qualitative agreement between the two methods, which shows that the automated method can be used on larger samples in future studies of how morphology depends on structural parameters and environment, thereby providing new constraints on the relative importance of various quenching mechanisms.

We have repeated the preceding analysis excluding the galaxies with $NUV-r\le4.5$ (described in \S\ref{smoothunsmooth}) and find that the results are qualitatively unchanged. This is because the these bluest galaxies are distributed across the range of each of our four size parameters, so that while the absolute numbers change, the overall trends stay the same.

\section{Summary}\label{summary}
We present a novel visual-based morphological classification scheme for red sequence galaxies based on SDSS images that is aimed at distinguishing between bulges and early type discs. Our interest in morphological information is motivated by the theoretical notion that mergers destroy stellar discs and create bulges. Future studies of bulge/disc frequencies using the new classification scheme may therefore provide a new constraint on the importance of mergers vs. other formation pathways in quenching galaxies with a variety of structural parameters in different environments.

The major element of the visual scheme uses the sharpness of the outer light profile to assign a bulge class index $BC$: galaxies with diffuse outer profiles are bulges ($BC=1$), while galaxies with sharp outer boundaries are discs ($BC=3$) . In addition, five features (spiral arms, bars, clumps, rings, and/or dust) that may indicate the presence of a cold disc component are collected into a single $FEAT$ index. Two more indices ($DIST$ and $INT$) measure large-scale morphological irregularities that might be caused by interactions and mergers.

The eventual goal of measuring bulge/disc frequencies as functions of structural and environmental parameters will require more galaxies than can be classified visually. We therefore attempt to reproduce the visual scheme using a set of four machine-measured parameters. The resulting recipe maps concentration $C$ and axis ratio $b/a$ from SDSS together with smoothness $s2$ and bulge fraction $B/T$ from GIM2D onto our visual parameters $BC$ and $FEAT$ (Figure~\ref{flowchart}).

A final comparison between the visual and automated methods are given in Table~\ref{bulgedisc_clean}. If the visual classifications are regarded as \textquoteleft truth,' the automated method identifies bulges with 75 per cent completeness and 73 per cent purity, and it identifies discs with 83 per cent completeness and 70 per cent purity. Most errors are one class; only 10 per cent of visual $BC=1$ bulges are completely misclassified as automated discs, and only 8 per cent of visual $BC=3$ discs are completely misclassified as automated bulges. Eight per cent of the entire sample of 984 galaxies has an error of more than one class. Plausible explanations are given for many of these cases (\S\ref{disagreements}).

The distribution of the inclinations of fitted discs sheds light on possible biases in the visual classifications. In particular, it is common wisdom that discs are more easily missed in face-on systems, and the cos $i$ distribution for GIM2D fitted discs is used to test for missing discs in both methods. Apart from an excess in one bin near cos $i\sim0.2$ the cos $i$ distributions in both methods are quite flat. There may be a slight loss of the most face-on discs using the visual method, but the statistical significance of this result is low. 

The inclination distributions of visual smooth vs. unsmooth discs show a stronger bias. Discs with features ($FEAT\ne1$) tend to be strongly face-on, indicating that features are often visually lost in edge-on discs. Moreover, the fraction of face-on discs (cos $i>0.8$) that show features is high ($\sim72$ per cent) indicating that real early type discs \textit{almost always} contain bars, arms, clumps, rings, and/or dust. This is one of our major results. Since $FEAT$ is evaluated completely separately from $BC$, the high frequency of features in our face-on $BC=3$ discs lends further strong support to our main criterion for identifying discs based on the sharpness of their outer light profiles.

Since the goal of this project is to study quenched galaxies, it is important minimize contamination by star-forming objects. We seek to do this by using only parameters widely available for SDSS galaxies in order to keep future samples large. We demonstrate that roughly half of all star-forming contaminants can be identified using either the visual or automated methods. Specifically, if unsmooth discs are equated to star-forming galaxies and are removed using the automated method, the resulting sample of 709 galaxies contains only 67 (9.4 per cent) visual unsmooth discs and only 80 (11 per cent) UV-blue galaxies. The main limitation for both methods is the angular size of the galaxy. When galaxies are too small, they are difficult to classify by eye and the automated parameters become less reliable (\S\ref{appendix}).

The method has been developed and tested on a sample of relatively low-$z$ galaxies ($z\sim0.06$) from the SDSS and can readily be used on future samples of large ($R_{90}>14''$), low-$z$ galaxies from the SDSS. A very large sample can be obtained by relaxing some of the criteria used here, for example, no longer requiring that the galaxies fall in the GALEX DR2 footprint.

The method has not been tested at high $z$, though it is likely to be useful out to $z\sim1$, where many galaxies have similar morphologies to low-$z$ samples (e.g., \citealp{lot08}) and where images from the Hubble Space Telescope can provide similar \textit{physical} resolution scales as those obtained by SDSS for our sample. It is likely to break down beyond $z > 1$, where galaxy morphologies may become significantly different from those observed at low redshift (e.g., \citealp{gen08,van08}). Further testing of the sort done in this paper will need to be done to apply the method to samples that are significantly different from the one used here.

\S\ref{results} conducts a preliminary exploration the distributions of bulges and discs using four different measures of galaxy size. Agreement between visual and automated methods is improved by assigning half of the intermediate class randomly to the bulge and disc samples (called bulges+ and discs+ respectively). In agreement with previous results, we find that bulges+ are in general brighter and have larger $M_*$, $\sigma$, and $R_{50}$ than discs+. The agreement between methods is quite good, though the trends appear to be slightly stronger when using the visual classifications. Figure~\ref{fundplane} shows a first attempt to map the distribution of bulges and discs directly onto the Fundamental Plane. Trends with $\sigma$ may be stronger than with $R_{50}$ in the plane, but this will have be checked with larger samples.

We have presented a method to classify \textquoteleft red and dead' galaxies in the SDSS using automated parameters from the galaxy fitting package GIM2D and the photometric pipeline of SDSS. In contrast to earlier automated methods, which typically distinguish only between early and late type galaxies, our method classifies early type galaxies into bulge- and disc-dominated classes. The agreement between our visual and automated classifications is quite good, with $<10 $ per cent of our sample in complete disagreement. Furthermore, the distributions of the different morphologies in galaxy \textquoteleft size' are also reproduced reasonably well by the automated scheme, which can now be applied to a larger sample of SDSS galaxies to explore how morphology varies with structural parameters, environment, UV flux, and other galaxy properties.

\section*{Acknowledgements}
We thank the reviewer for many useful suggestions that improved the clarity and cogency of this work. MCC acknowledges support by NASA through the Spitzer Space Telescope Fellowship Program. This work was supported by NSF grants AST05 and AST08.

Funding for the Sloan Digital Sky Survey (SDSS) has been provided by the Alfred P. Sloan Foundation, the Participating Institutions, the National Aeronautics and Space Administration, the National Science Foundation, the U.S. Department of Energy, the Japanese Monbukagakusho, and the Max Planck Society. The SDSS Web site is http://www.sdss.org/.

The SDSS is managed by the Astrophysical Research Consortium (ARC) for the Participating Institutions. The Participating Institutions are The University of Chicago, Fermilab, the Institute for Advanced Study, the Japan Participation Group, The Johns Hopkins University, the Korean Scientist Group, Los Alamos National Laboratory, the Max-Planck-Institute for Astronomy (MPIA), the Max-Planck-Institute for Astrophysics (MPA), New Mexico State University, University of Pittsburgh, University of Portsmouth, Princeton University, the United States Naval Observatory, and the University of Washington.

\appendix

\section{More on the Validity of the GIM2D and SDSS Parameters}\label{appendix}
Figure~\ref{bf_conc} shows that the most bulge-dominated galaxies also tend to be the most concentrated. We note, however, several features that bear more examination.

\subsection{GIM2D Bulge Fraction $B/T$}\label{validity_bf}
The first notable feature in Figure~\ref{bf_conc} is the outliers with high values of $C$. We expect that the most concentrated objects ($C\sim3.5$) should also have the highest bulge fractions $B/T$. Instead, we see that they have intermediate bulge fractions $0.5 < B/T < 0.6$ with a slight downturn in $C$ at higher values of $B/T$. This occurs because the GIM2D fits that we are using require that the bulge component have a S\'ersic index $n=4$, which is not always a good fit, as many giant ellipticals are known to have $n>4$. These galaxies may have more light in the wings than an $n=4$ profile can account for, and a disc is added to improve the fit. The fraction of light attributed to the disc is thus higher, and the bulge fraction is underestimated.

We have investigated this effect by analyzing analogous GIM2D $r$-band fits where $n$ is treated as a free parameter. These fits result in a $B/T$-$C$ relation that is not peaked at intermediate values of $B/T$ as in Figure~\ref{bf_conc} but rises continuously and gradually to $B/T=1$. While the value of $B/T$ changes for many galaxies, most of the galaxies for which $B/T$ changes significantly have $B/T>0.6$ and do not fall near the boundary. For visual discs with high $B/T$ ($BC=3$, $B/T > 0.6$), the distribution of S\'ersic indices peaks between $n=3$ and $n=4$, which indicates that assuming an $n=4$ fit is reasonable for these galaxies. Overall, the results of our automated classifications using the two fits are similar and the choice of $n=4$ versus floating-$n$ fits has no significant effect on the results presented here. 

\begin{figure}
\includegraphics[width=0.49\textwidth]{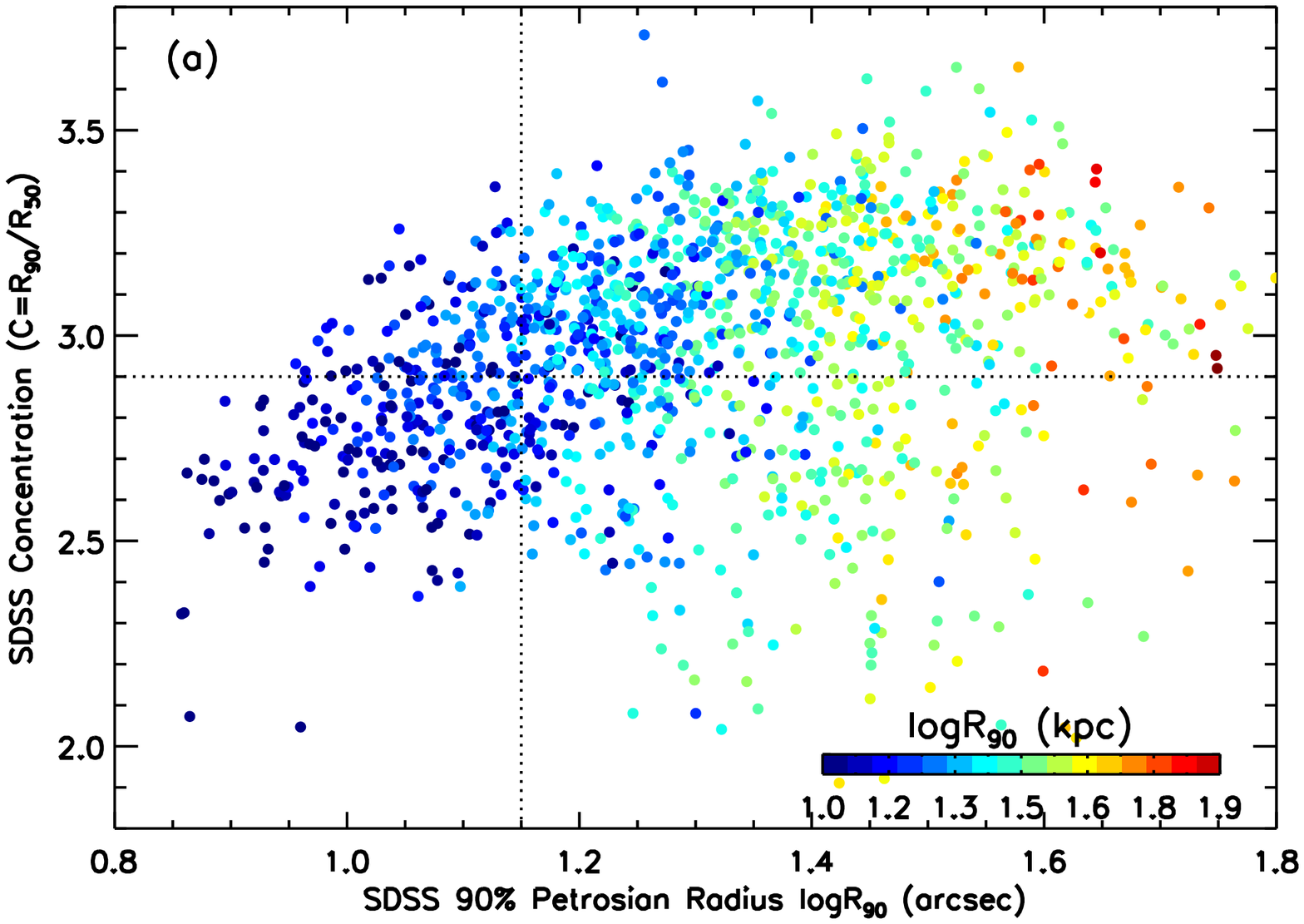}
\includegraphics[width=0.49\textwidth]{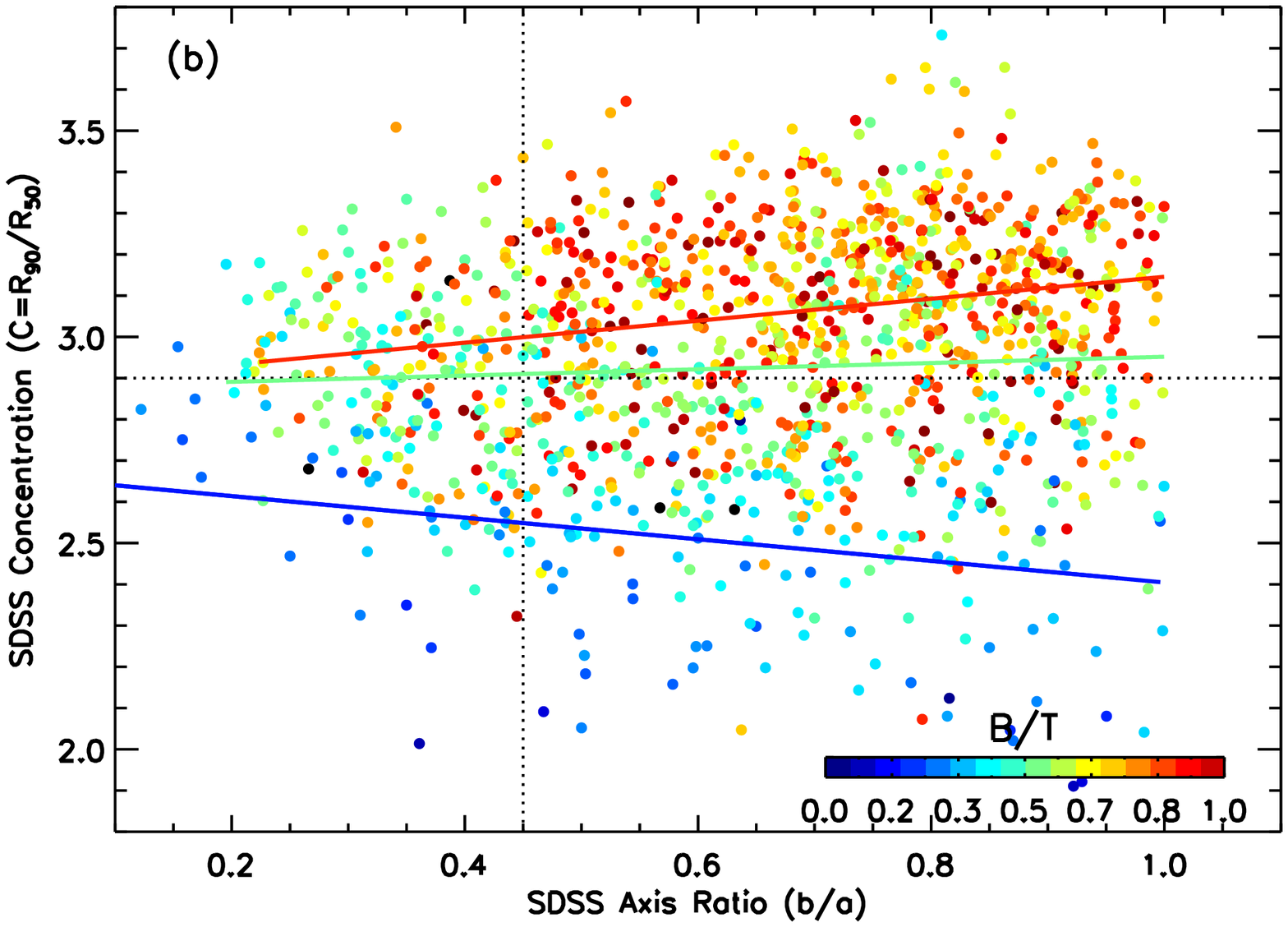}
\caption{Distributions of SDSS concentration $C$. (a) SDSS 90 per cent Petrosian Radius vs. SDSS concentration for 1295 galaxies, with color indicating \textit{absolute physical} size. The vertical dotted line marks our adopted \textquoteleft smallness' cut at log $R_{90}=1.15$ (\S\ref{visualscheme}), while the horizontal dotted line marks the $C=2.9$ cut used in the automated scheme in Figure~\ref{bf_conc}. For small values of $R_{90}$, there appears to be a ceiling to the value of $C$; as the apparent size of the galaxy decreases, the maximum value of $C$ decreases as well. The smallness cut screens out most such objects. (b) SDSS axis ratio vs. SDSS concentration for 1295 galaxies, with color indicating GIM2D bulge fraction $B/T$. Three linear fits to different $B/T$ bins are shown. The vertical dotted line indicates the $b/a=0.45$ cut used in the automated scheme in Figure~\ref{bf_abr}. The measured value of $C$ appears to behaved differently as a function of $b/a$ for bulges and discs. In particular, edge-on discs (low $b/a$) have higher concentration $C$ that face-on discs (high $b/a$) with comparable $B/T$. This effect has also been noted by \citet{yam05} and \citet{bai08}. We conclude from these panels that the value of $C$ is least reliable for the smallest and most elongated galaxies.}
\label{conc}
\end{figure}

\subsection{SDSS Concentration $C$}\label{validity_c}
The second notable feature in Figure~\ref{bf_conc} is the group of galaxies that appears to fall below the main $B/T$-$C$ relation, in the automated intermediate region. These generally prove to be small and faint. Figure~\ref{conc}(a) shows SDSS concentration $C$ as a function of SDSS 90 per cent Petrosian radius $R_{90}$ (i.e., galaxy \textit{apparent} size) for the initial sample of 1295 objects with color indicating \textit{absolute physical} size. We use the full sample of 1295 because we are interested in how the apparent size of the galaxy affects the value $C$, especially for the small galaxies removed by the \textquoteleft smallness' cut at log $R_{90}=1.15$ in \S\ref{visualscheme}. This cut is indicated by the vertical dotted line, while the adopted boundary between automated bulges and intermediates at $C=2.9$ is indicated by the horizontal dotted line. For large apparent size, the low $C$ galaxies tend to be large face-on discs, based on the visual classifications in Figure~\ref{bf_conc}. For small apparent size, there appears to be a ceiling to the value of $C$; as the apparent size of the galaxy decreases, the maximum value of $C$ decreases as well, which is logical, as seeing smooths out the central peak in high-concentration galaxies. Furthermore, this maximum is not a strong function of the galaxy's \textit{absolute physical} size.

The last notable feature in Figure~\ref{bf_conc} is the large number of visual discs with rather high values of $C$, even at low values of $B/T$. We find that these are often galaxies with elongated appearances (i.e., edge-on discs). Figure~\ref{conc}(b) shows SDSS concentration $C$ as a function of SDSS axis ratio $b/a$ for the same sample, with color indicating GIM2D bulge fraction $B/T$. The vertical dotted line indicates the value of $b/a$ below which objects are classified as automated discs in Figure~\ref{bf_abr}, while the horizontal dotted line indicates the value of $C$ below which objects are classified as automated intermediates in Figure~\ref{bf_conc}. 

Two trends are seen here. First, the lack of galaxies with high $C$ and low $b/a$ is consistent with the idea that bulge-dominated galaxies (high $B/T$, red) are both round and centrally concentrated. Second, the measured value of $C$ changes systematically for discs (low $B/T$, blue): edge-on discs (low $b/a$) have higher concentration $C$ than face-on discs (high $b/a$) with comparable $B/T$. This is illustrated with linear fits to galaxies with different values of $B/T$ (indicated by color). This effect, as noted in work by \citet{yam05} and \citet{bai08}, occurs because $R_{90}$ is measured using circular apertures. We conclude from Figure~\ref{conc} that the value of $C$ is least reliable for both the smallest and the most elongated galaxies.

Our automated classifications are unlikely to be affected greatly by either effect. The former effect appears to be strongest for galaxies log $R_{90}\le1.15$, all of which were eliminated from the final sample.  There may be a slight effect on small galaxies near the boundary, which may have concentrations $\sim0.2$ less than their \textquoteleft true' values. As for the latter effect, the most elongated galaxies ($b/a\le0.45$) are already selected to be in the automated disc sample based on their axis ratios and will not contaminate the automated bulge sample even though they have high values of $C$.

\bibliographystyle{apj}
\bibliography{cheng2010_arxiv}

\end{document}